\documentclass[aps,pre,preprint,superscriptaddress]{revtex4-1}
\usepackage{amsmath}
\usepackage{graphicx}
\usepackage{url}
\usepackage{hyperref}

\graphicspath{{Figures/}}

\begin{document}

%\preprint{}

\title{Simulation study of the electrical tunneling network conductivity of suspensions of hard spherocylinders}


\author{Arshia Atashpendar}
\email{arshia.atashpendar@physik.uni-freiburg.de}
\affiliation{Institute of Physics, University of Freiburg, Germany}

\author{Sarthak Arora}

\affiliation{Department of Physics, IIT Guwahati, India}

\author{Alexander D. Rahm}

\affiliation{Mathematics Research Unit, Faculty of Science, Technology and Communication, University of Luxembourg}

\author{Tanja Schilling}

\affiliation{Institute of Physics, University of Freiburg, Germany}


\date{\today}

\begin{abstract}
Using Monte Carlo simulations, we investigate the electrical conductivity of networks of hard rods with aspect ratios $10$ and $20$ as a function of the volume fraction for two tunneling conductance models. For a simple, orientationally independent tunneling model, we observe non-monotonic behaviour of the bulk conductivity as a function of volume fraction at the isotropic-nematic transition. However, this effect is lost if one allows for anisotropic tunneling. The relative conductivity enhancement increases exponentially with volume fraction in the nematic phase. Moreover, we observe that the orientational ordering of the rods in the nematic phase induces an anisotropy in the conductivity, i.e.~enhanced values in the direction of the nematic director field. We also compute the mesh number of the Kirchhoff network, which turns out to be a simple alternative to the computationally expensive conductivity of large systems in order to get a qualitative estimate. 
\end{abstract}

%\keywords{}

\maketitle

\section{Introduction}
%\label{}
At the percolation transition, clusters of connected objects merge to form one system-spanning network. This transition was first studied in the context of the flow of water through porous rock. By now, due to the technological relevance of transport processes in disordered media, in general, as well as the mathematical properties of percolation transition, percolation has become a topic of interest in a wide range of different research fields \cite{torquato:2002, aharony:2003}.
In this article, we focus on the formation of networks of electrically conducting fibres, as they would be used e.g.~in a composite material that is designed to undergo a transition from an insulating state to a conductive state. These types of composites have been studied extensively for almost 50 years in experiment, simulation and theory. A complete literature overview is hence beyond the scope of this paper. Instead, we refer the reader to a selection of relevant articles in the context of the present work: For reviews on conductive composites see refs.~\cite{Mutiso2014, Thomassin13, park:2013, sanjines2011electrical, schilling:2009, Huang02}, for theoretical approaches to connectivity percolation in rods see refs.~\cite{balberg:1984, balberg:1985, balberg:1986, munson-mcgee:1991, leung:1991, drwenski2017connectedness, chatterjee:2000, wang:2003, yi:2004, Otten2009, chatterjee2015overview, kale2015tunneling, Dixit16, Schilling15, Meyer15, mecke:2002}, for simulation refs.~\cite{pike:1974, lee:1988, foygel:2005, rahatekar2005mesoscale, berhan:2007, Skvor07a, akagawa:2007, ambrosetti:2010, schilling:2007, Miller09c, mutiso:2012, Mutiso2013} and for experiments refs.~\cite{coleman:1998, sandler:1999, barrau:2003, vigolo:2005, lyons2008relationship, Nirmalraj:2012, Cattin14, majidian2017role, Grossiord2008, Kyrylyuk2011, kumar2016tuning}. 

In suspension, anisotropic particles form liquid crystalline phases. In particular, upon increasing the concentration of rod-like particles, they undergo a transition from an isotropic phase to an orientationally ordered, but positionally disordered phase (called the "nematic" phase)\cite{onsager1949effects, p1995physics}. In general, the electrical conductivity of a suspension of aligned particles is different from that of an isotropic system. The effect of alignment on networks of conductive fillers has been studied in experiment \cite{ackermann2016effect,wan2017highly,wang2008effects, choi2003enhancement, du2005effect} and in theory and simulation \cite{finner2018continuum, kale2016effect, zheng:2005, Otten12, lebovka2018anisotropy}. However, these studies considered alignment by means of an external field such as e.g.~shear and not alignment due to a phase transition that is intrinsic to the system. To our knowledge, there is no systematic study of the behaviour of the conductivity across the isotropic-nematic (I-N) phase transition yet. It is thus the aim of our study to fill this gap.

\section{Models and Simulations}

We have modeled only the filler particles and neglected the host environment of the nanocomposite. We used hard spherocylinders ("rods") with a cylindrical part of length $L$ and diameter $D$, capped on both ends with hemispheres of diameter $D$. The rods are 'coated' by a spherocylindrical penetrable contact shell of diameter $\lambda,$ which serves as a means of establishing a geometric connectivity criterion between pairs of rods, where if two rods have overlapping contact shells they are considered to be a connected pair. By extension, a cluster can be formed by a contiguous sequence of such pairwise connected rods. 

We have generated configurations of hard rods using canonical Monte Carlo simulations (NVT-MC). The simulations have been carried out for systems of aspect ratio $L/D = 20$ and $10$, using a non-cubic\footnote{where the director of the nematic phase was roughly parallel to the elongated z-direction} and a cubic simulation box of dimensions $L_{x,y}=3L, L_z=4L,$ and $L_{x,y,z}=6L,$ respectively. The particle numbers ranged from $N=1000$ to $5000$ for $L/D=20$ and $N=6000$ to $8000$ for $L/D=10$ in order to cover a range of volume fractions that span configurations in both the isotropic and the nematic phase. The volume fraction is defined as $\eta = N v_{\rm core}/V,$ with $N$ being the number of rods in suspension, $V$ volume of the simulation box, and $v_{\rm core}=\pi D^2 (L/4+D/6)$ denotes the volume of one rod. 

We work with two different contact shell diameters for the case of $L/D=20$, namely, $\lambda=1.1 D$ and $\lambda = 1.2 D$. The former is small enough such that irrespective of the volume fraction we simulated, the suspensions never exhibited a percolating cluster (i.e., they appear as a sparse collection of isolated clusters), while conversely, for the latter choice of $\lambda,$ the rods were always in a percolating state (i.e., the suspension is mostly comprised of a large connected cluster spanning the simulation box). Despite the fact that the electrical connections between the rods will be subsequently modeled by tunneling processes, which decay continuously with the inter-particle distance and thus without a sharp cut-off, the motivation behind the aforementioned $\lambda$ choices and their corresponding geometric percolation states is the fact that they provide us with two contrasting connectivity states for the same liquid state of the suspension. Ultimately, they exemplify two cases of short and long tunneling decay lengths, which we will touch upon in more detail subsequently. However, in the case of aspect ratio $10,$ we will only consider one contact shell value, $\lambda=1.1 D,$ which is large enough to ensure a similar percolating state for all simulated volume fractions. 

As a scalar order parameter to distinguish between the isotropic and nematic phase, we use $S_2$, the maximum eigenvalue of the orientation tensor $Q$ as given by:

\begin{equation}\label{eq:s2}
    Q_{ij} = \frac{1}{2N}\sum_{\alpha=1}^N (3v_i^{\alpha}v_j^{\alpha}-\delta_{ij})
\end{equation}
where $v_i^{\alpha}$ and $v_j^{\alpha}$ are the ith and jth components of the normalized orientation vector of rod $\alpha,$ respectively, and $\delta_{ij}$ is the Kronecker delta.

For the conductance $g_{ij}$ between a given pair of rods, we use the commonly adopted model of single-electron tunneling as the dominant process for electron transfer in composite systems dispersed with nano-sized conductive filler particles\cite{ambrosetti:2010}. The general form of the model is given by a simple exponential decay function $g:$

\begin{equation}\label{eq:expdecay}
g(d_{ij})= g_0 \exp{(-2d_{ij}/\xi)}
\end{equation}
where $d_{ij}$ denotes the shortest distance between the surfaces of rods $i$ and $j$, and $\xi$ is the tunneling decay length (which is usually taken to be in the order of magnitude of a few nanometers and in our simulations it is simply taken as difference between the penetrable shell diameter $\lambda$ and the impenetrable rod diameter $D$). Conventionally, we set the conductance between two rods whose surfaces are touching to be equal to unity which resolves the pre-factor in \eqref{eq:expdecay} into being unity as well: $g(0)=1.$
It is important to discuss the range of validity of Eq.~\eqref{eq:expdecay}, which is sometimes also referred to as \textit{normal tunneling.} Briefly, the limit of validity of normal tunneling as the dominant process of electron transport between nano-sized conductive filler particles, is directly related to the cross-section area of the tunneling junction between them, namely, the narrower the junction, the higher the probability for events of single uncorrelated electrons tunneling through the junction barrier. The cross-section area depends primarily on the shape of the electrodes, which in our case are spherocylindrical. Furthermore, for the general working temperature of nanocomposites, namely $T<500K,$ normal tunneling is expected to remain valid. However, given that hard systems are athermal, together with the fact that the area of the junction is then dependent only on particle geometry and relative orientation with a neighbouring particle, the former validity criterion can be translated into one on the aspect ratio, $L/D \ll 200$. Thus, for our purposes the choices of $L/D=10,20$ remain reasonable if we are to adopt Eq.~\eqref{eq:expdecay} in order to describe pairwise conductances. For an in-depth discussion on the validity of normal tunneling in systems of elongated particles, we refer the reader to the work of Sherman et al.~\cite{sherman1983electron}. 

Next to Eq.~\eqref{eq:expdecay}, we also applied a second conductance model, which deviates from Eq.~\eqref{eq:expdecay} by its explicit incorporation of the relative orientation between two given rods in order to account for the cross section of the tunneling junction between the pair. More precisely, G. Nigro and C. Grimaldi~\cite{nigro2014impact} have recalculated the matrix element for electron tunneling between rod-like particles and have shown that in general the tunneling between two parallel rods (e.g., in the nematic phase) is more probable by a factor of $L/\sqrt{\xi D}$ compared to two rods perpendicular to one another. Moreover, they show that for disordered phases (e.g., in the isotropic phase) the orientationally dependent contributions to the conductance become negligible once an averaging over all solid angles is performed. However, for strongly oriented phases of suspension of rod-like particles, the enhanced tunneling between parallel rods is expected to have a significant influence on the conductivity of the network of rods. In order to  draw a comparison with the results that are obtained according to Eq.~\ref{eq:expdecay}, we will adopt a simplified form of the anisotropic tunneling conductance, which is obtained with the underlying assumption that the centres of mass of two connected rods are close to their line of shortest distance\cite{chatterjee2015tunneling}:

\begin{equation}\label{eq:anisoconduc}
g(d_{ij},\gamma_{ij})/g_0 = \frac{\exp{(-2d_{ij}/\xi})}{\frac{2\pi D \xi}{L^2}+\sin^2(\gamma_{ij})}
\end{equation}
where the newly introduced variable $\gamma_{ij}$ denotes the angle between the long axes of the rods.
We note that, our comparison merely aims at probing the effects on the bulk conductivity resulting from a simple deviation from the common model Eq.~\ref{eq:expdecay}, one that would account in an explicit manner for the relative orientation when estimating the tunneling conductance between two rods. Therefore, we take one of the simplest forms of the corrections derived by G. Nigro and C. Grimaldi~\cite{nigro2014impact}, which offers the added advantage of maintaining the numerical computations rather simple. Moreover considering that on the one hand, we have only chosen to work with short aspect ratio rods, namely of $20$ and $10,$ contained in simulation boxes of dimensions $4$ and $6$ times larger than $L$ respectively, and that on the other hand, only small tunneling decay lengths ($10-20\%$ of rod diameter) are considered, the assumptions behind Eq.~\ref{eq:anisoconduc} remain reasonable for our purposes. 

\begin{figure}
    \centering
    \includegraphics[width=0.3\linewidth]{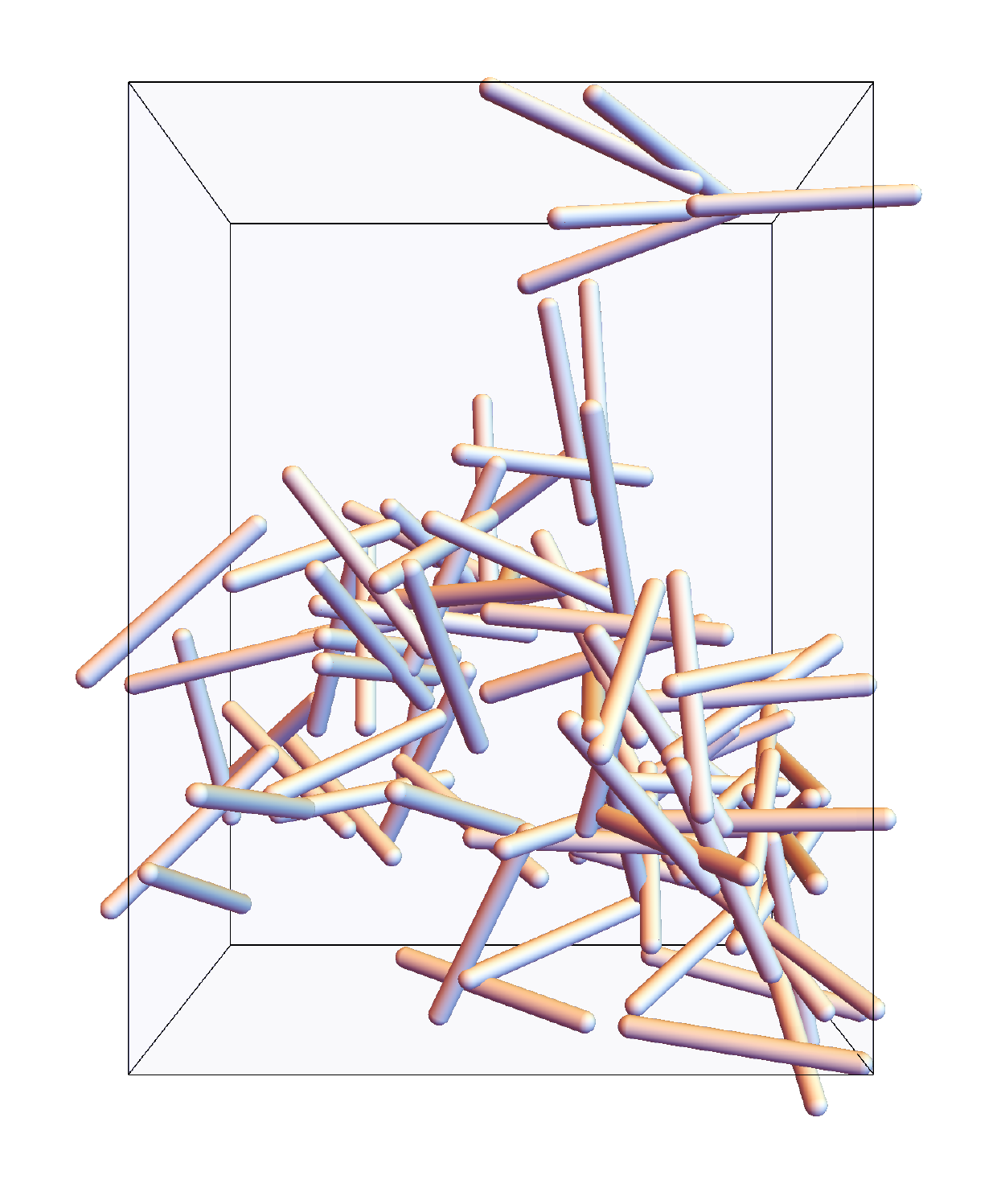}
    \includegraphics[width=0.3\linewidth]{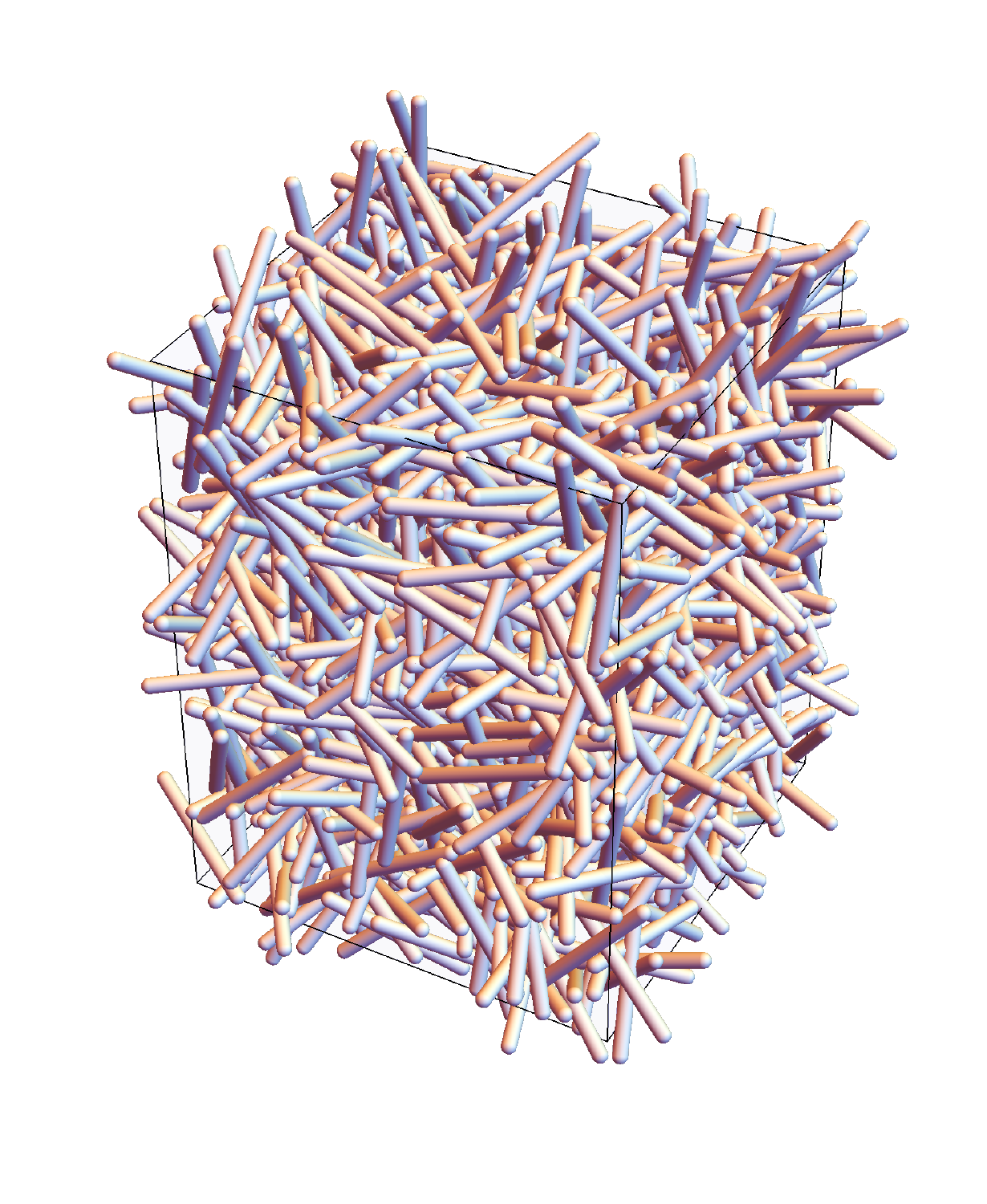}
    \includegraphics[width=0.3\linewidth]{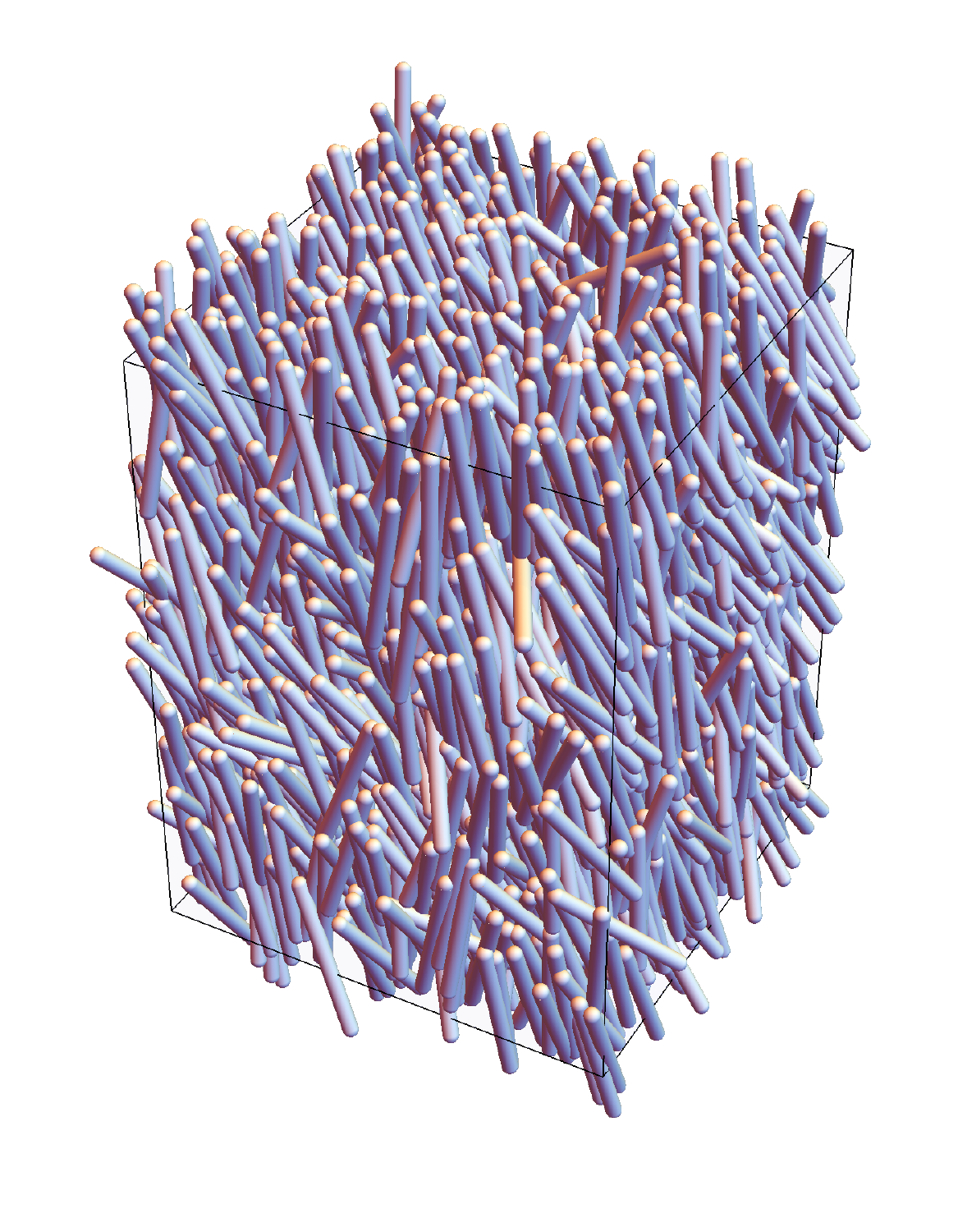}
    \caption{Snapshots of percolating clusters at different volume fractions.}
    \label{fig:Clu_snapshots}
\end{figure}

Lastly, we briefly discuss our methodology for estimating the network conductivity $\sigma,$ a calculation which is only performed on the largest cluster of a given simulated suspension. That is, at a given volume fraction and irrespective of whether the chosen contact shell diameter is below or above the percolation threshold, we first reduce the corresponding configuration (equilibrium suspension of hard rods) to its largest cluster, where clusters are defined according to our geometric connectivity criterion of overlapping penetrable shells of diameter $\lambda.$ In other words, the reduction to clusters ensures that we perform the conductivity calculations on subsets of rods where for any given rod, there is at least one neighbour such that the shortest distance between their long axes is $\le \xi + D = \lambda.$ Then, having identified the largest cluster, we proceed to the construction of the tunneling resistor network. In Fig.~\ref{fig:Clu_snapshots} examples of percolating largest clusters in the isotropic and nematic phase are visualized. The resistor (conductance) network is built by assigning a node to each rod and a resistor between all pairs of nodes in the system (comprised of the largest cluster) whose value is given by the inverse of their corresponding conductance either given by Eq.~\eqref{eq:expdecay} or \eqref{eq:anisoconduc}. For measuring the conductivity, two electrodes are placed on the most distant nodes except in the study of the conductivity anisotropy in the nematic phase where we additionally restrict their connecting axis to be once parallel and once orthogonal to the common director field of the rods. 

In order to solve the Kirchhoff equations (i.e., to obtain the equivalent resistance or equivalently the overall conductivity) of the corresponding tunneling-based resistor network, we measure the equivalent resistance between the two electrode-nodes by performing a DC simulation on the network in Qucs~\cite{margraf2017qucs}. More precisely, in order to obtain the equivalent resistance between the electrode-nodes, a constant current source of $1 A$ is placed between them and the voltage across the two nodes is measured during the simulation. As a second method for comparison, in a handful of cases we also apply an exact numerical decimation using the Star-Mesh transform, which entails applying Rosen's formula~\cite{middendorf1956analysis} until all nodes except the electrode-nodes are eliminated. In order to reduce the computational efforts due to the large number of tunneling bonds in the network ($\propto N(N-1),$ with $N$ here denoting the number of rods in the largest cluster), we remove the tunneling bonds between rods that are sufficiently apart, meaning their tunneling conductance is negligibly small to be relevant in the estimation of the bulk conductivity. Thus, their removal would not be in conflict with any of our previous considerations, and as long as the introduced cut-off is at least equal or greater than $\lambda,$ the resulting simplified network shares the same connectivity properties, such as the nearest neighbour distances, and in particular, the same percolation state. Therefore, given the rapid exponential decay of the conductances and the known decay length scales, we introduced an artificial cut-off distance of $\lambda^2/D$ which considerably simplified the tunneling network and in turn reduced the computational times needed for the decimation routine. The final conductivity value for each volume fraction is averaged over an ensemble of $200$ independent largest clusters. It is important to point out that by reducing the configuration of the rods to their largest cluster in addition to the aforementioned cut-off serving to lower computational costs, the resulting conductivity estimate will consequently represent a lower bound for the corresponding equilibrium suspension at a given volume fraction.

In an independent calculation, we also counted the number of meshes of the simplified Kirchhoff network: We recorded the edge-vertex-incidence matrix of the network corresponding to the largest cluster considered as an undirected graph, and then computed the mesh number from the dimension of its kernel, i.e., by subtracting its rank from the number of resistors (size of the graph). In order to be consistent with the conductivity calculations, we have again used the same cut-off when setting the edges of the graph, i.e., it is interpreted as a graph where a bond is placed between any two nodes separated by a distance $<\lambda^2/D.$ It is important to note that the mesh calculations and results are completely independent of the choice of conductance model, and instead, depend only on the geometric connectivity properties of the network formed by the rods.

\section{Results}

\begin{figure}
    \centering
    \includegraphics[width=0.98\linewidth]{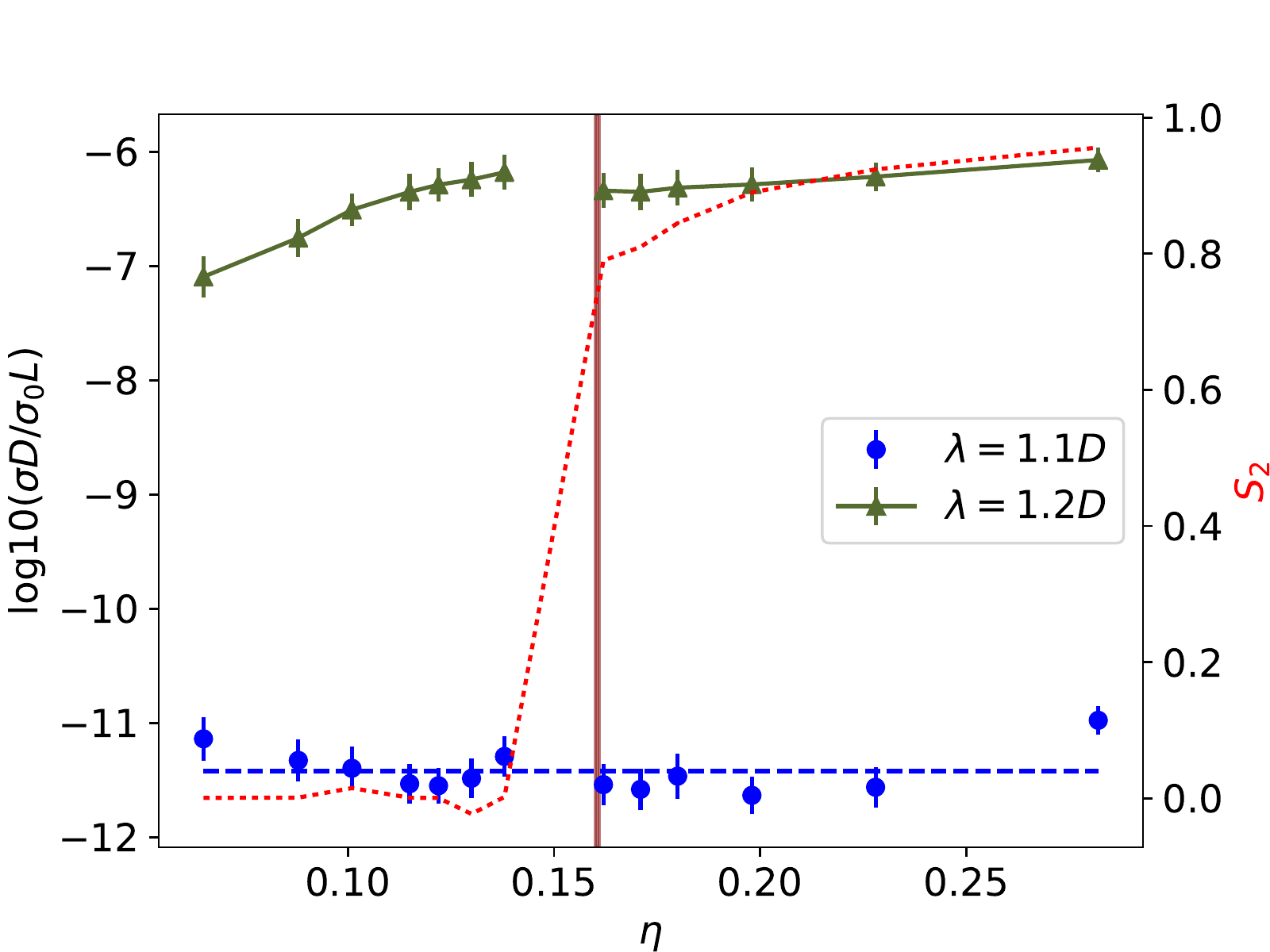}
    \caption{Conductivity (in $\log$ base $10$) for the normal tunneling model Eq.~\ref{eq:expdecay} as a function of volume fraction for two different contact shell diameters, $\lambda=1.1 D$ (blue symbols, mean value as a dashed line), and $\lambda=1.2 D$ (green symbols, solid line). In addition, on the second Y-axis the mean nematic order parameter ($S_2,$ red, dotted line) values of the entire system (not just the connected clusters) is plotted, and the vertical solid (brown) line indicates the coexistence volume fraction in the nematic phase for $L/D=20.$ (colour online)}
    \label{fig:sigma_expdecay_diffLambda}
\end{figure}

We start by discussing the bulk conductivity ($\sigma$) as a function of volume fraction ($\eta$) for values in the isotropic and the nematic phase and for two contact shell diameters, $\lambda=1.1\,D, 1.2\, D$ and $L/D=20$. In Figure~\ref{fig:sigma_expdecay_diffLambda} we show results for normal tunneling (eqn.~\ref{eq:expdecay}). The conductivity increases with volume fraction for the percolating case (for $\lambda=1.2\,D$), while below the percolation threshold ($\lambda=1.1\,D$) we observe a constant and negligibly small value for $\sigma$. In the latter case, the conductivity of the nano-composite would correspond to that of the polymeric host material alone. 

\begin{figure}
    \centering
    \includegraphics[width=0.75\linewidth]{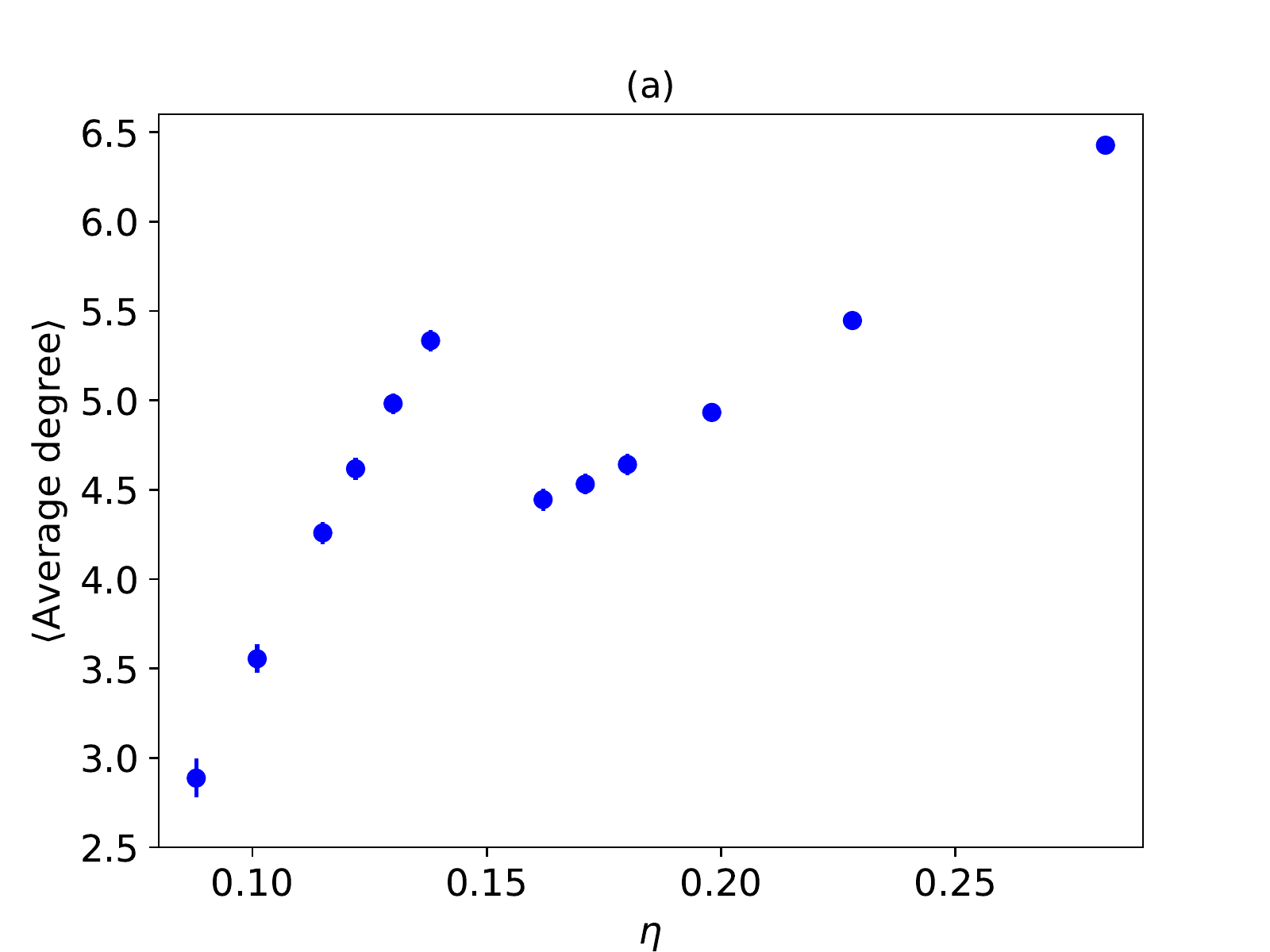}
    \includegraphics[width=0.75\linewidth]{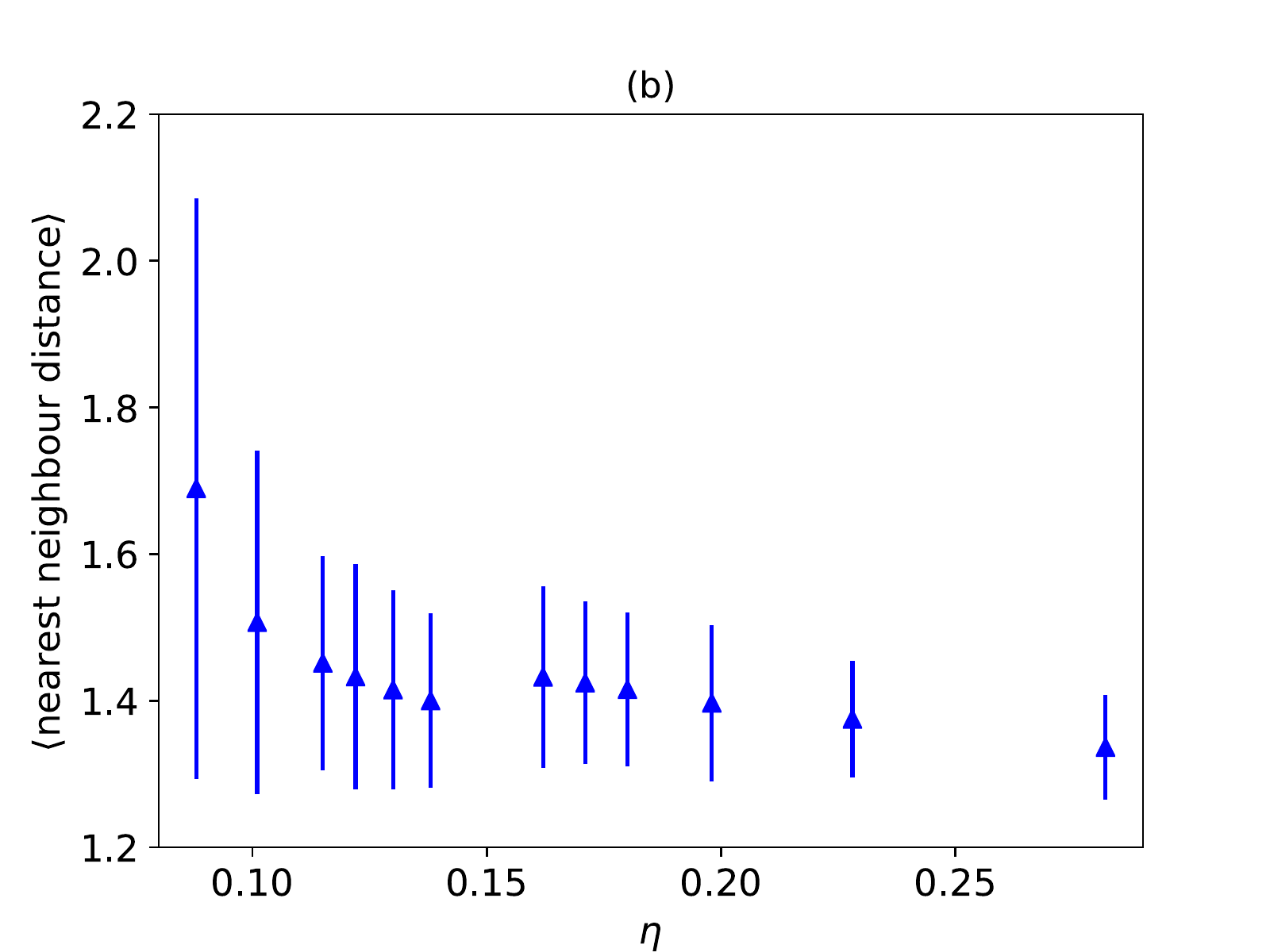}
    \caption{(a): The average degree of a rod as a function of volume fraction for $L/D=20$ and $\lambda=1.2D.$ The average degree is simply calculated as twice the size of the edge set divided by the number of nodes in a given network. (b): The mean distance to a rod's nearest neighbour in the largest cluster as a function of volume fraction. In both plots, the error bars are obtained by averaging over all independent realisations of largest clusters per volume fraction.}
    \label{fig:AveDeg_MeanDist_AR20}
\end{figure}

In Fig.~\ref{fig:sigma_expdecay_diffLambda} we also show the nematic order parameter $S_2$ (red dotted line, second y-axis) of the entire system. 
At the onset of the nematic phase (vertical solid line) and for $\lambda=1.2D$, despite the increase in volume fraction, we observe a clear drop in $\sigma$ relative to the last value observed in the isotropic phase. After the drop at the onset, $\sigma$ increases again with $\eta$, but less strongly than in the isotropic phase \footnote{On a related note, for an analysis based on the effective medium approximation (EMA) and critical path approximation (CPA) of the dependence of $\sigma$ on the interplay of volume fraction and orientational ordering, we refer the reader to the work of A. P. Chatterjee and C. Grimaldi~\cite{chatterjee2015tunneling}.}.
This non-monotonic behaviour is due to the increase in orientational order at the phase transition, which has the effect of reducing the excluded volume between neighbouring rods. Consequently, the latter translates into a weakened connectivity of the resulting network (here interpreted as a graph where the edges are assigned geometrically based on our cut-off distance criterion). The said weakening has been successfully captured in the plots of Fig.~\ref{fig:AveDeg_MeanDist_AR20}, in terms of the behaviour of the average degree and the nearest neighbour distance of a rod as a function of volume fraction. More precisely, in plot (a) of Fig.~\ref{fig:AveDeg_MeanDist_AR20}, the average degree, simply taken as twice the number of edges divided by the number of nodes (rods), exhibits the same non-monotonic behaviour as $\sigma$, where at the I-N transition the number of connected neighbouring rods can be clearly seen to become smaller before increasing again. Furthermore, in plot (b) of Fig.~\ref{fig:AveDeg_MeanDist_AR20}, the mean nearest neighbour distance of a rod in a cluster is shown to decrease with increasing volume fraction, but the observed increase at the I-N transition suggests a reduction in the largest conductance contribution per rod in the network. Therefore, we observe a consistent correlation between the weakened-geometrically-interpreted connectivity properties of the network of rods and its corresponding tunneling conductivity.

\begin{figure}
    \centering
    \includegraphics[width=0.75\linewidth]{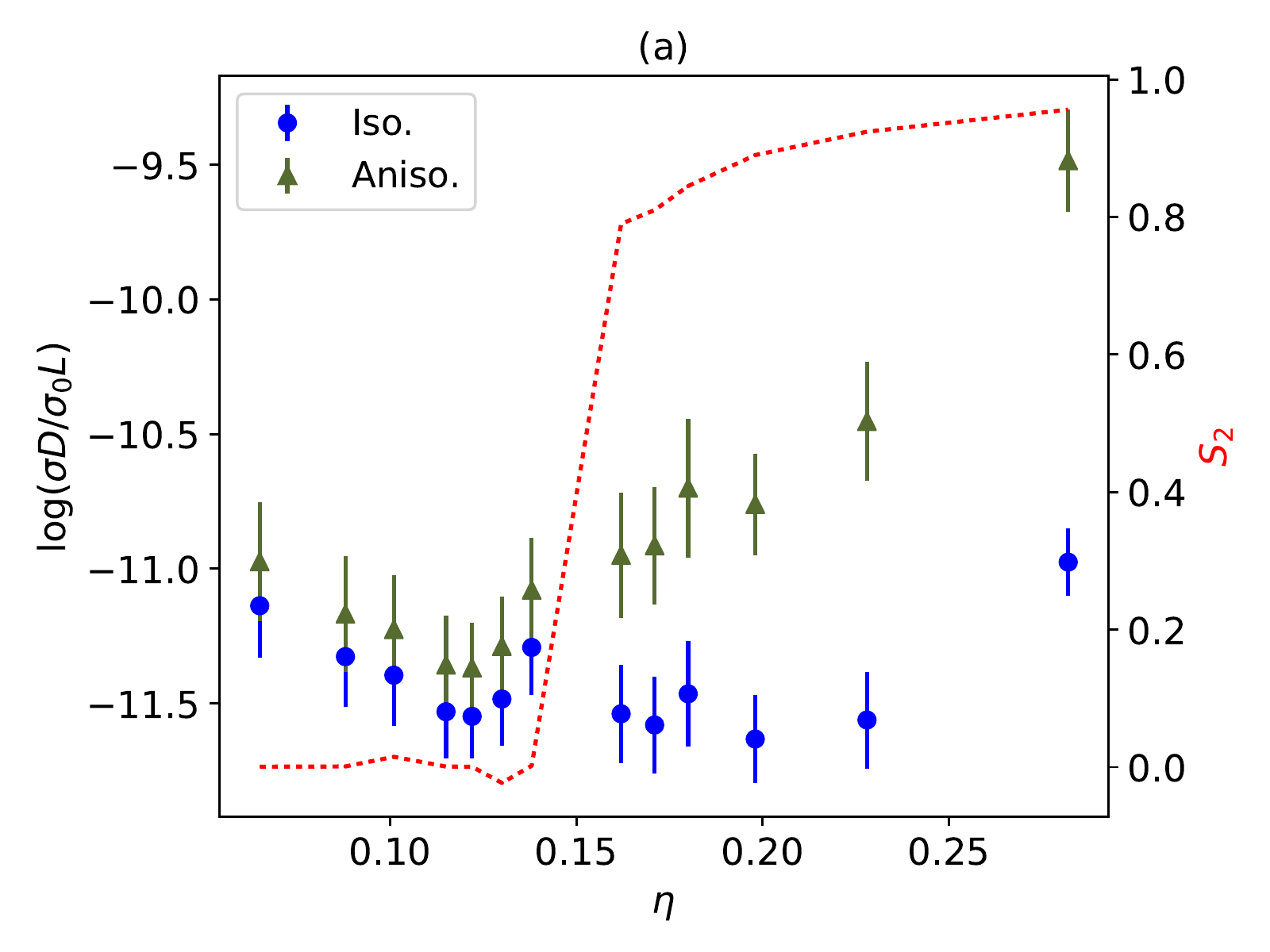}
    \includegraphics[width=0.75\linewidth]{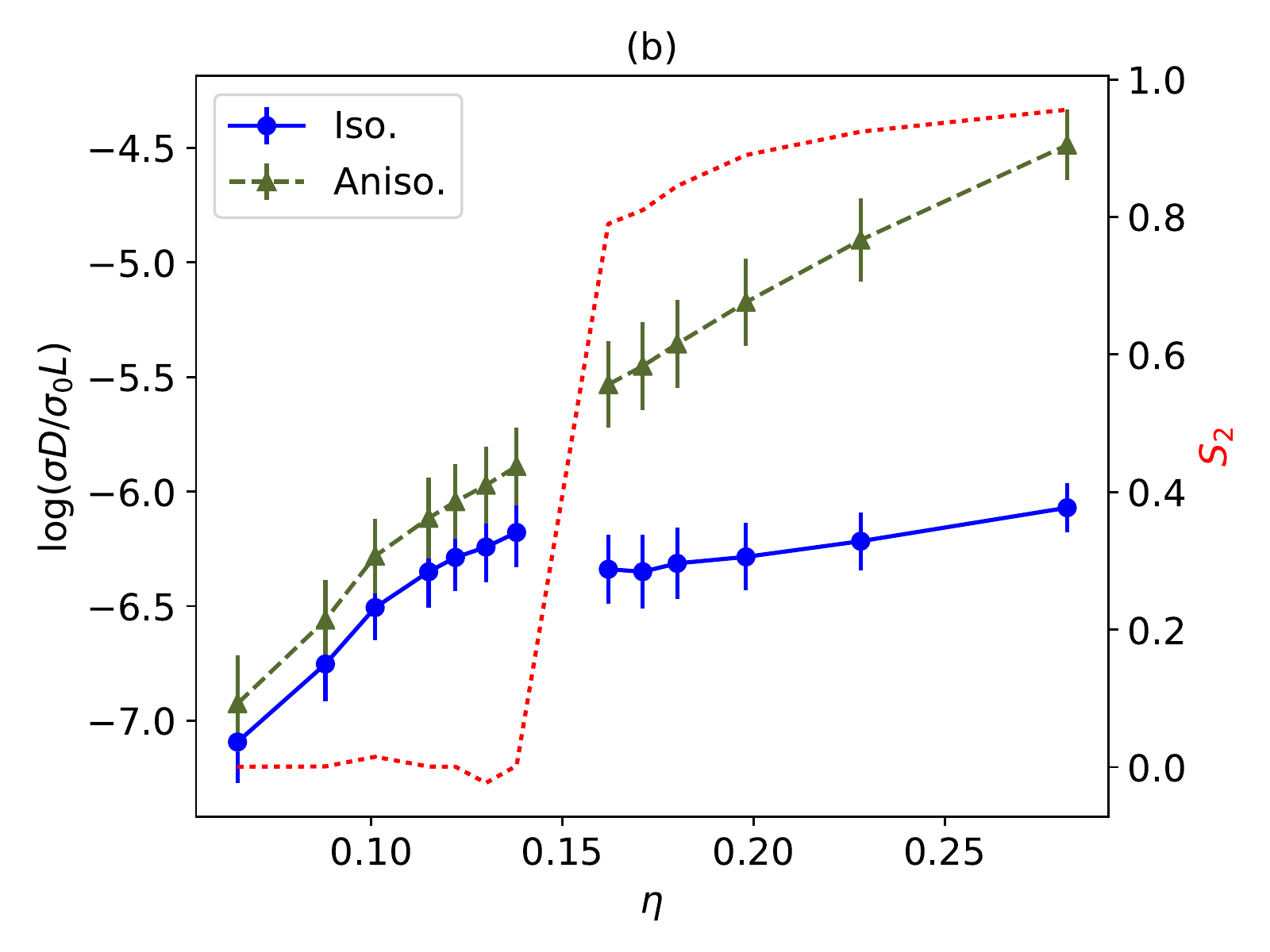}
    \caption{Comparison of the dependence of $\sigma$ on $\eta$ for the two tunneling conductance models as given in Eq.~\ref{eq:expdecay} (labelled \textit{Iso.}) and Eq.~\ref{eq:anisoconduc} (labelled \textit{Aniso.}), and for $L/D=20.$ (a): $\lambda=1.1D$, (b): $\lambda=1.2 D.$ The nematic order parameter $S_2$ is shown on the second Y-axis in each case.}
    \label{fig:sigma_iso_aniso}
\end{figure}

\begin{figure}
    \centering
    \includegraphics[width=0.98\linewidth]{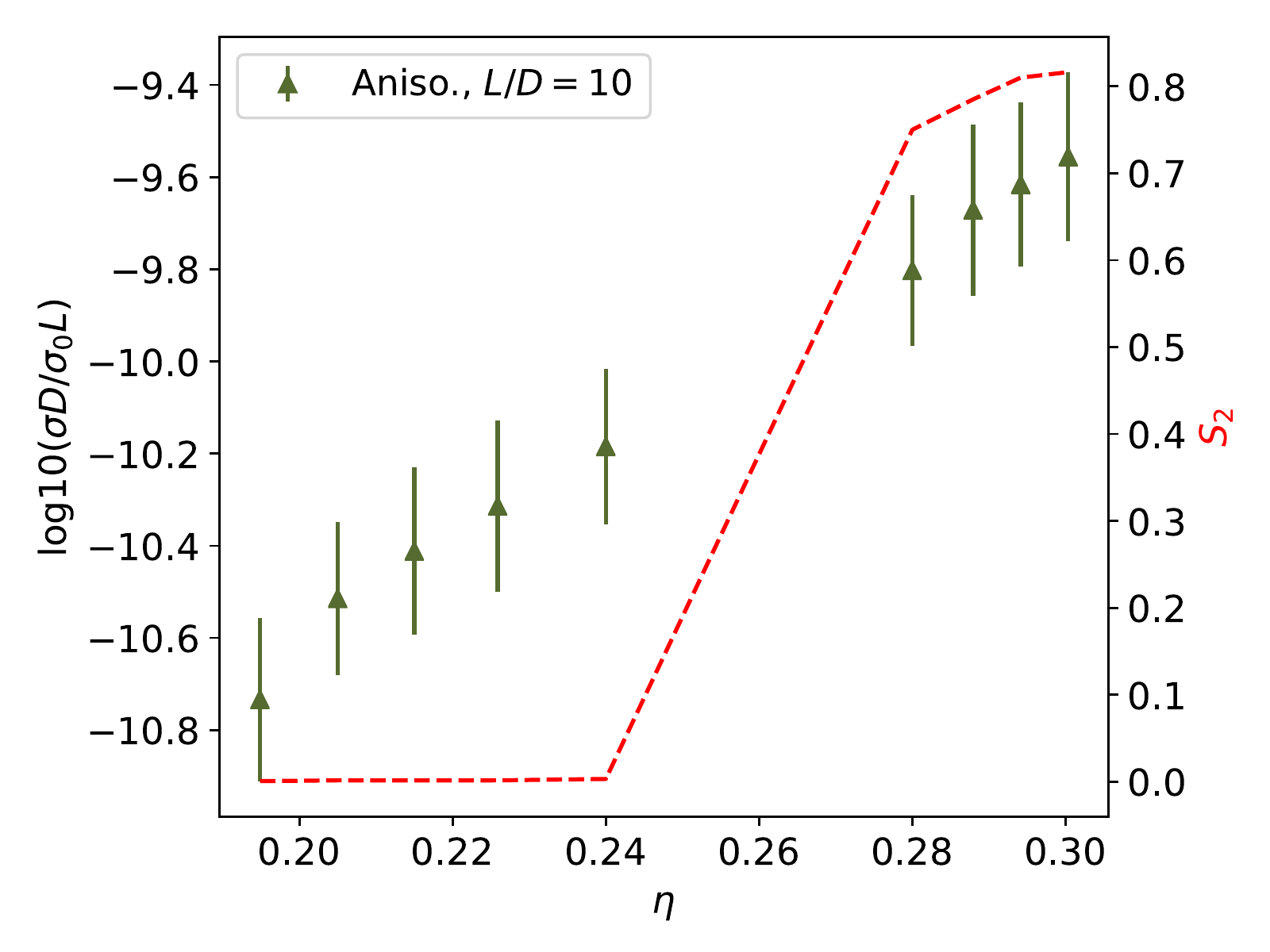}
    \caption{Conductivity for the anisotropic tunnel model Eq.~\ref{eq:anisoconduc} as a function of volume fraction, for $L/D=10$ and $\lambda=1.1 D,$ which is above the percolation threshold for the entire of range of chosen volume fractions here. The mean nematic order parameter ($S_2$) is plotted on the second Y-axis (red dashed).}
    \label{fig:sigma_aniso_AR10}
\end{figure}

Next we compare the dependence of $\sigma$ on $\eta$ for the two different tunneling conductance models given in Eqs.~\ref{eq:expdecay}~and~\ref{eq:anisoconduc}. In Fig.~\ref{fig:sigma_iso_aniso}, both below and above the percolation threshold, we observe that the anisotropic tunneling model enhances the conductivity, in particular in the nematic phase. The non-monotonic behaviour of the conductivity across the transition is lost for the more realistic, anisotropic model. That is, the conductance enhancement between aligned rods inherent to the model dominates the previously characterized effect of the weakened connectivity that the network undergoes at the I-N transition. In Fig.~\ref{fig:sigma_aniso_AR10}, the latter behaviour is also observed for $L/D=10$, based on the model Eq.~\ref{eq:anisoconduc} and using $\lambda=1.1 D,$ which is above the percolation threshold.

The enhancement of the conductivity in nano-composites across the I-N transition of the filler particles has been previously theoretically predicted~\cite{zheng:2005} in the context of polymeric fillers of very high aspect ratio ($L/D=100$ and above) under shear flow. Zheng et al showed that the relative conductivity enhancement \footnote{More precisely, first an effective expression for the electrical conductivity tensor is derived, then the largest relative enhancement is defined in terms of the maximum eigenvalue of the effective conductivity tensor after the inclusion of conductive filler rods. For a clear account of all details, we refer the reader to the complete work of Zheng et al.~\cite{zheng:2005}.}, which is the relative conductivity change in the polymeric bulk after and before the dispersion of conductive filler particles, grows linearly with volume fraction both in the isotropic and nematic stable regions. 

\begin{figure}
    \centering
    \includegraphics[width=0.98\linewidth]{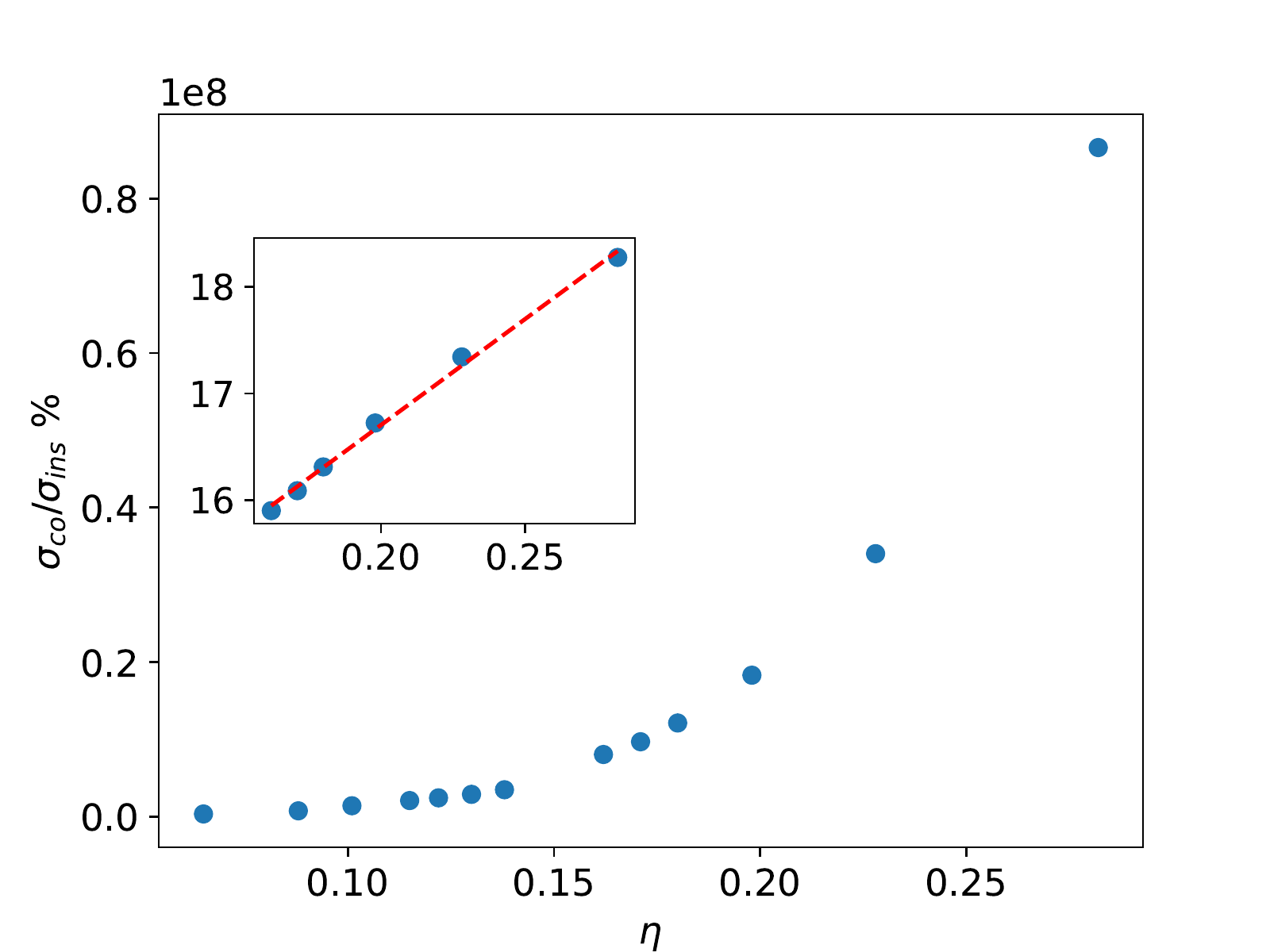}
    \caption{The relative conductivity enhancement in percentages for the anisotropic tunneling model as a function of volume fraction, which is calculated based on the data shown in Fig.~\ref{fig:sigma_iso_aniso}. $\sigma_{\rm co}$ denotes $\sigma$ in the conductive state, i.e., in the presence of percolating clusters ($\lambda=1.2 D$), and $\sigma_{ins}$ denotes the mean conductivity value in the absence of percolating clusters ($\lambda=1.1 D$). The inset plot shows $\ln{\sigma_{\rm co}/\sigma_{\rm ins}}$ as a function of $\eta$ restricted to the nematic phase, with the linear fit shown in dashed.}
    \label{fig:enhancement}
\end{figure}

In order to draw a closer comparison in addition to the general qualitative agreement, we set the conductivity in the host insulating material as $\sigma_{\rm ins}=\langle \sigma \rangle_{\lambda=1.1}$ and similarly the conductivity in the conductive state to $\sigma_{\rm co}=\sigma_{\lambda=1.2},$ i.e., $\sigma$ in the presence of percolating clusters. With the latter choices, we define the relative conductivity enhancement as the ratio $\epsilon=\sigma_{\rm co}/\sigma_{\rm ins}.$ In Fig.~\ref{fig:enhancement}, the relative enhancement based on the anisotropic conductance model is plotted as a function of $\eta.$ We clearly observe that the relative enhancement is well described in the nematic phase by an exponential function, $\epsilon(\eta)\approx \exp(\alpha \eta),$ with $\alpha>0.$, i.e.~we observe exponential rather than linear scaling. In the inset plot of Fig.~\ref{fig:enhancement}, a linear fit applied to $\ln{(\epsilon)}$ restricted to the volume fraction range in the nematic phase yields $\alpha=19.9\pm 0.9.$

We note that due to the fact that in this comparison, neither the aspect ratios, nor the considered conductivities in the host insulating environment (more precisely the ratios of $\sigma$ after and before dispersion) are the same, a direct quantitative comparison with ref.~\cite{zheng:2005} remains difficult to perform. 

\begin{figure}
    \centering
    \includegraphics[width=0.98\linewidth]{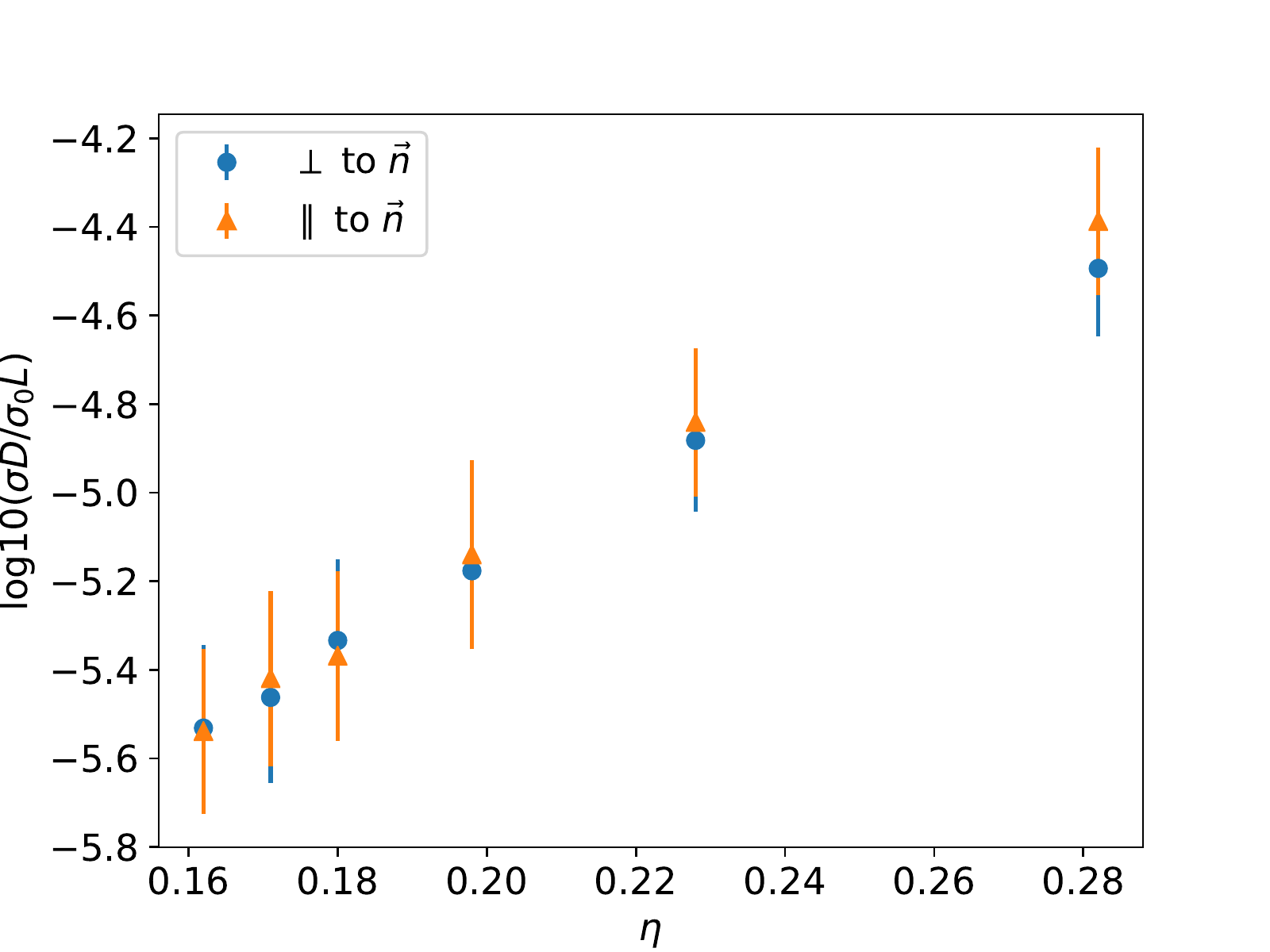}
    \caption{Conductivity in the nematic phase for $L/D=20, \lambda=1.2D,$ using the model Eq.~\ref{eq:anisoconduc}, as a function of volume fraction. $\sigma$ measured along the director $\Vec{n}$ of the rods (orange, triangular), compared to $\sigma$ measured in the orthogonal direction to $\Vec{n}$ (blue, round).}
    \label{fig:sigma_DiffElectrodes}
\end{figure}

In particular, for the aligned phase of the rods in the bulk, having already studied the enhancement of $\sigma$ compared to the disordered phase, next, we briefly discuss our observations on the induced anisotropy of $\sigma$. Using the model Eq.~\ref{eq:anisoconduc} and $\lambda=1.2D$,  we compare $\sigma$ as measured in parallel and orthogonal directions to the common director field $\vec{n}$ of the rods in the nematic phase. The comparison is shown in Fig.~\ref{fig:sigma_DiffElectrodes}, where near the nematic coexistence volume fraction we observe on average the same behaviour for $\sigma$ along the two directions with respect to $\vec{n}.$ On the other hand, for volume fractions past this region, i.e., with increasing orientational ordering of the rods, we notice a clear separation in trends where $\sigma$ values measured along the director field become increasingly larger with $S_2.$ This reflects the fact that the increased orientational ordering in the stable region of the nematic phase does introduce anisotropy (favoured direction) in the bulk conductivity, which has been a consistent observation in various experimental works~\cite{ackermann2016effect,wan2017highly,wang2008effects, choi2003enhancement, du2005effect}.

\begin{figure}
    \centering
    \includegraphics[width=0.98\linewidth]{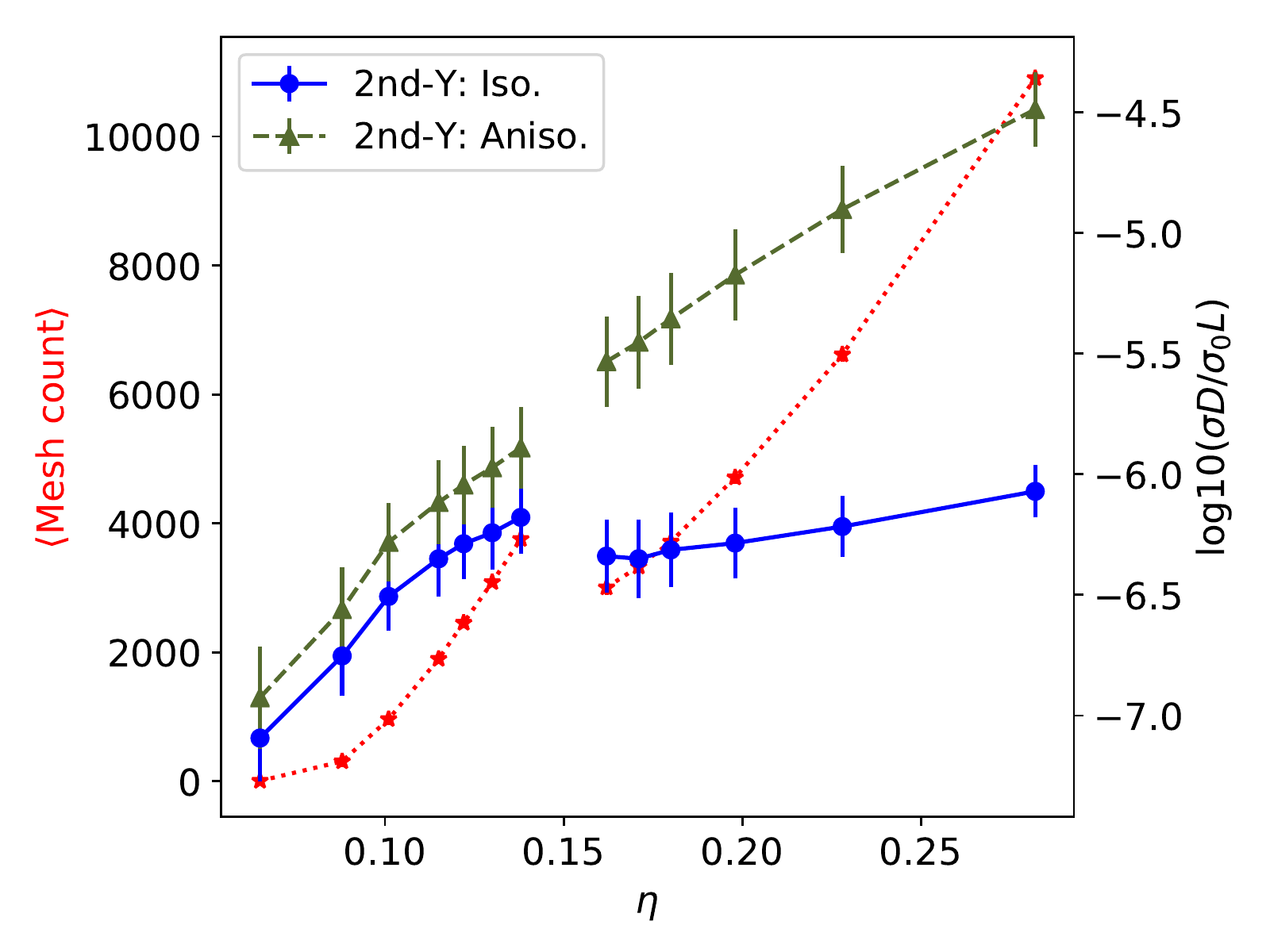}
    \caption{Mean mesh numbers (red, star and dotted) and conductivity (2nd Y-axis, solid and dashed curves) as a function of $\eta,$ for $L/D=20$ and $\lambda=1.2 D.$}
    \label{fig:1_2_mesh_sigma}
\end{figure}

Finally, we compare the trend in $\sigma$ with the mean number of meshes corresponding to the largest clusters of rods (interpreted as graphs with edges assigned according to our introduced cut-off distance) as a function of $\eta$.  This comparison is shown in Fig.~\ref{fig:1_2_mesh_sigma}. Restricted to either phases of the rods, we observe that the mean mesh number follows the same increasing trend with $\eta$ as we have observed for $\sigma.$ However, there are clear qualitative differences in their functional form. More importantly, given that the mesh count abstracts away from the actual conductance model and depends solely on the geometric connectivity state of the rods, it exhibits the same non-monotonic behaviour in $\eta$ at the I-N transition that we have previously observed for the different graph properties (mean degree and nearest neighbour distance) and also for $\sigma$ in the normal tunneling case. This offers the potential for a simpler and computationally less expensive route to extracting the trend of $\sigma$ restricted to a given phase of the rods, without explicitly solving the Kirchhoff equations of the tunneling network. 

\section{Conclusions}
We have studied the conductivity of networks of hard spherocylinders in the isotropic and the nematic phase. For the widely-used normal tunneling model we observe non-monotonic behaviour of the conductivity at the phase transition. However, for a more realistic anisotropic tunneling model, this effect is lost. Furthermore, we observe that the conductivity in the nematic phase is isotropic close to the phase transition, but becomes anisotropic at higher volume fractions. We also analyzed the number of meshes in the geometrically interpreted network of rods and found it to be a very rough, but fast to compute, estimator for the conductivity.

\begin{acknowledgments}
The authors acknowledge support by the state of Baden-W\"urttemberg through bwHPC
and the German Research Foundation (DFG) through grant no INST 39/963-1 FUGG. Alexander D. Rahm is supported by Universit\'e du Luxembourg through Gabor Wiese's AMFOR grant.
\end{acknowledgments}

% If in two-column mode, this environment will change to single-column
% format so that long equations can be displayed. Use
%\begin{widetext}
%\end{widetext}

\clearpage

% \bibliography{Bibliography}

\begin{thebibliography}{67}%
\makeatletter
\providecommand \@ifxundefined [1]{%
 \@ifx{#1\undefined}
}%
\providecommand \@ifnum [1]{%
 \ifnum #1\expandafter \@firstoftwo
 \else \expandafter \@secondoftwo
 \fi
}%
\providecommand \@ifx [1]{%
 \ifx #1\expandafter \@firstoftwo
 \else \expandafter \@secondoftwo
 \fi
}%
\providecommand \natexlab [1]{#1}%
\providecommand \enquote  [1]{``#1''}%
\providecommand \bibnamefont  [1]{#1}%
\providecommand \bibfnamefont [1]{#1}%
\providecommand \citenamefont [1]{#1}%
\providecommand \href@noop [0]{\@secondoftwo}%
\providecommand \href [0]{\begingroup \@sanitize@url \@href}%
\providecommand \@href[1]{\@@startlink{#1}\@@href}%
\providecommand \@@href[1]{\endgroup#1\@@endlink}%
\providecommand \@sanitize@url [0]{\catcode `\\12\catcode `\$12\catcode
  `\&12\catcode `\#12\catcode `\^12\catcode `\_12\catcode `\%12\relax}%
\providecommand \@@startlink[1]{}%
\providecommand \@@endlink[0]{}%
\providecommand \url  [0]{\begingroup\@sanitize@url \@url }%
\providecommand \@url [1]{\endgroup\@href {#1}{\urlprefix }}%
\providecommand \urlprefix  [0]{URL }%
\providecommand \Eprint [0]{\href }%
\providecommand \doibase [0]{http://dx.doi.org/}%
\providecommand \selectlanguage [0]{\@gobble}%
\providecommand \bibinfo  [0]{\@secondoftwo}%
\providecommand \bibfield  [0]{\@secondoftwo}%
\providecommand \translation [1]{[#1]}%
\providecommand \BibitemOpen [0]{}%
\providecommand \bibitemStop [0]{}%
\providecommand \bibitemNoStop [0]{.\EOS\space}%
\providecommand \EOS [0]{\spacefactor3000\relax}%
\providecommand \BibitemShut  [1]{\csname bibitem#1\endcsname}%
\let\auto@bib@innerbib\@empty
%</preamble>
\bibitem [{\citenamefont {Torquato}(2002)}]{torquato:2002}%
  \BibitemOpen
  \bibfield  {author} {\bibinfo {author} {\bibfnamefont {S.}~\bibnamefont
  {Torquato}},\ }\href@noop {} {\emph {\bibinfo {title} {Random heterogeneous
  materials: microstructure and macroscopic properties}}},\ Vol.~\bibinfo
  {volume} {16}\ (\bibinfo  {publisher} {Springer},\ \bibinfo {year}
  {2002})\BibitemShut {NoStop}%
\bibitem [{\citenamefont {Aharony}\ and\ \citenamefont
  {Stauffer}(2003)}]{aharony:2003}%
  \BibitemOpen
  \bibfield  {author} {\bibinfo {author} {\bibfnamefont {A.}~\bibnamefont
  {Aharony}}\ and\ \bibinfo {author} {\bibfnamefont {D.}~\bibnamefont
  {Stauffer}},\ }\href@noop {} {\emph {\bibinfo {title} {Introduction to
  percolation theory}}}\ (\bibinfo  {publisher} {Taylor \& Francis},\ \bibinfo
  {year} {2003})\BibitemShut {NoStop}%
\bibitem [{\citenamefont {Mutiso}\ and\ \citenamefont
  {Winey}(2014)}]{Mutiso2014}%
  \BibitemOpen
  \bibfield  {author} {\bibinfo {author} {\bibfnamefont {R.~M.}\ \bibnamefont
  {Mutiso}}\ and\ \bibinfo {author} {\bibfnamefont {K.~I.}\ \bibnamefont
  {Winey}},\ }\href {\doibase 10.1016/j.progpolymsci.2014.06.002} {\enquote
  {\bibinfo {title} {{Electrical properties of polymer nanocomposites
  containing rod-like nanofillers}},}\ } (\bibinfo {year} {2014})\BibitemShut
  {NoStop}%
\bibitem [{\citenamefont {Thomassin}\ \emph {et~al.}(2013)\citenamefont
  {Thomassin}, \citenamefont {Jerome}, \citenamefont {Pardoen}, \citenamefont
  {Bailly}, \citenamefont {Huynen},\ and\ \citenamefont
  {Detrembleur}}]{Thomassin13}%
  \BibitemOpen
  \bibfield  {author} {\bibinfo {author} {\bibfnamefont {J.-M.}\ \bibnamefont
  {Thomassin}}, \bibinfo {author} {\bibfnamefont {C.}~\bibnamefont {Jerome}},
  \bibinfo {author} {\bibfnamefont {T.}~\bibnamefont {Pardoen}}, \bibinfo
  {author} {\bibfnamefont {C.}~\bibnamefont {Bailly}}, \bibinfo {author}
  {\bibfnamefont {I.}~\bibnamefont {Huynen}}, \ and\ \bibinfo {author}
  {\bibfnamefont {C.}~\bibnamefont {Detrembleur}},\ }\href {\doibase
  10.1016/j.mser.2013.06.001} {\bibfield  {journal} {\bibinfo  {journal}
  {MATERIALS SCIENCE \& ENGINEERING R-REPORTS}\ }\textbf {\bibinfo {volume}
  {74}},\ \bibinfo {pages} {211} (\bibinfo {year} {2013})}\BibitemShut
  {NoStop}%
\bibitem [{\citenamefont {Park}\ \emph {et~al.}(2013)\citenamefont {Park},
  \citenamefont {Vosguerichian},\ and\ \citenamefont {Bao}}]{park:2013}%
  \BibitemOpen
  \bibfield  {author} {\bibinfo {author} {\bibfnamefont {S.}~\bibnamefont
  {Park}}, \bibinfo {author} {\bibfnamefont {M.}~\bibnamefont {Vosguerichian}},
  \ and\ \bibinfo {author} {\bibfnamefont {Z.}~\bibnamefont {Bao}},\
  }\href@noop {} {\bibfield  {journal} {\bibinfo  {journal} {Nanoscale}\
  }\textbf {\bibinfo {volume} {5}},\ \bibinfo {pages} {1727} (\bibinfo {year}
  {2013})}\BibitemShut {NoStop}%
\bibitem [{\citenamefont {Sanjin{\'e}s}\ \emph {et~al.}(2011)\citenamefont
  {Sanjin{\'e}s}, \citenamefont {Abad}, \citenamefont {V{\^a}ju}, \citenamefont
  {Smajda}, \citenamefont {Mioni{\'c}},\ and\ \citenamefont
  {Magrez}}]{sanjines2011electrical}%
  \BibitemOpen
  \bibfield  {author} {\bibinfo {author} {\bibfnamefont {R.}~\bibnamefont
  {Sanjin{\'e}s}}, \bibinfo {author} {\bibfnamefont {M.~D.}\ \bibnamefont
  {Abad}}, \bibinfo {author} {\bibfnamefont {C.}~\bibnamefont {V{\^a}ju}},
  \bibinfo {author} {\bibfnamefont {R.}~\bibnamefont {Smajda}}, \bibinfo
  {author} {\bibfnamefont {M.}~\bibnamefont {Mioni{\'c}}}, \ and\ \bibinfo
  {author} {\bibfnamefont {A.}~\bibnamefont {Magrez}},\ }\href@noop {}
  {\bibfield  {journal} {\bibinfo  {journal} {Surface and coatings technology}\
  }\textbf {\bibinfo {volume} {206}},\ \bibinfo {pages} {727} (\bibinfo {year}
  {2011})}\BibitemShut {NoStop}%
\bibitem [{\citenamefont {Schilling}\ \emph {et~al.}(2009)\citenamefont
  {Schilling}, \citenamefont {Jungblut},\ and\ \citenamefont
  {Miller}}]{schilling:2009}%
  \BibitemOpen
  \bibfield  {author} {\bibinfo {author} {\bibfnamefont {T.}~\bibnamefont
  {Schilling}}, \bibinfo {author} {\bibfnamefont {S.}~\bibnamefont {Jungblut}},
  \ and\ \bibinfo {author} {\bibfnamefont {M.~A.}\ \bibnamefont {Miller}},\
  }\href@noop {} {\bibfield  {journal} {\bibinfo  {journal} {Handbook of
  Nanophysics}\ } (\bibinfo {year} {2009})}\BibitemShut {NoStop}%
\bibitem [{\citenamefont {Huang}(2002)}]{Huang02}%
  \BibitemOpen
  \bibfield  {author} {\bibinfo {author} {\bibfnamefont {J.}~\bibnamefont
  {Huang}},\ }\href {\doibase 10.1002/adv.10025} {\bibfield  {journal}
  {\bibinfo  {journal} {ADVANCES IN POLYMER TECHNOLOGY}\ }\textbf {\bibinfo
  {volume} {21}},\ \bibinfo {pages} {299} (\bibinfo {year} {2002})}\BibitemShut
  {NoStop}%
\bibitem [{\citenamefont {Balberg}\ \emph {et~al.}(1984)\citenamefont
  {Balberg}, \citenamefont {Anderson}, \citenamefont {Alexander},\ and\
  \citenamefont {Wagner}}]{balberg:1984}%
  \BibitemOpen
  \bibfield  {author} {\bibinfo {author} {\bibfnamefont {I.}~\bibnamefont
  {Balberg}}, \bibinfo {author} {\bibfnamefont {C.~H.}\ \bibnamefont
  {Anderson}}, \bibinfo {author} {\bibfnamefont {S.}~\bibnamefont {Alexander}},
  \ and\ \bibinfo {author} {\bibfnamefont {N.}~\bibnamefont {Wagner}},\
  }\href@noop {} {\bibfield  {journal} {\bibinfo  {journal} {Phys. Rev. B}\
  }\textbf {\bibinfo {volume} {30}},\ \bibinfo {pages} {3933} (\bibinfo {year}
  {1984})}\BibitemShut {NoStop}%
\bibitem [{\citenamefont {Balberg}(1985)}]{balberg:1985}%
  \BibitemOpen
  \bibfield  {author} {\bibinfo {author} {\bibfnamefont {I.}~\bibnamefont
  {Balberg}},\ }\href@noop {} {\bibfield  {journal} {\bibinfo  {journal} {Phys.
  Rev. B}\ }\textbf {\bibinfo {volume} {31}},\ \bibinfo {pages} {4053}
  (\bibinfo {year} {1985})}\BibitemShut {NoStop}%
\bibitem [{\citenamefont {Balberg}(1986)}]{balberg:1986}%
  \BibitemOpen
  \bibfield  {author} {\bibinfo {author} {\bibfnamefont {I.}~\bibnamefont
  {Balberg}},\ }\href@noop {} {\bibfield  {journal} {\bibinfo  {journal} {Phys.
  Rev. B}\ }\textbf {\bibinfo {volume} {33}},\ \bibinfo {pages} {3618}
  (\bibinfo {year} {1986})}\BibitemShut {NoStop}%
\bibitem [{\citenamefont {Munson-McGee}(1991)}]{munson-mcgee:1991}%
  \BibitemOpen
  \bibfield  {author} {\bibinfo {author} {\bibfnamefont {S.~H.}\ \bibnamefont
  {Munson-McGee}},\ }\href@noop {} {\bibfield  {journal} {\bibinfo  {journal}
  {Phys. Rev. B}\ }\textbf {\bibinfo {volume} {43}},\ \bibinfo {pages} {3331}
  (\bibinfo {year} {1991})}\BibitemShut {NoStop}%
\bibitem [{\citenamefont {Leung}\ and\ \citenamefont
  {Chandler}(1991)}]{leung:1991}%
  \BibitemOpen
  \bibfield  {author} {\bibinfo {author} {\bibfnamefont {K.}~\bibnamefont
  {Leung}}\ and\ \bibinfo {author} {\bibfnamefont {D.}~\bibnamefont
  {Chandler}},\ }\href@noop {} {\bibfield  {journal} {\bibinfo  {journal} {J.
  Stat. Phys.}\ }\textbf {\bibinfo {volume} {63}},\ \bibinfo {pages} {837}
  (\bibinfo {year} {1991})}\BibitemShut {NoStop}%
\bibitem [{\citenamefont {Drwenski}\ \emph {et~al.}(2017)\citenamefont
  {Drwenski}, \citenamefont {Dussi}, \citenamefont {Dijkstra}, \citenamefont
  {van Roij},\ and\ \citenamefont {van~der
  Schoot}}]{drwenski2017connectedness}%
  \BibitemOpen
  \bibfield  {author} {\bibinfo {author} {\bibfnamefont {T.}~\bibnamefont
  {Drwenski}}, \bibinfo {author} {\bibfnamefont {S.}~\bibnamefont {Dussi}},
  \bibinfo {author} {\bibfnamefont {M.}~\bibnamefont {Dijkstra}}, \bibinfo
  {author} {\bibfnamefont {R.}~\bibnamefont {van Roij}}, \ and\ \bibinfo
  {author} {\bibfnamefont {P.}~\bibnamefont {van~der Schoot}},\ }\href@noop {}
  {\bibfield  {journal} {\bibinfo  {journal} {The Journal of chemical physics}\
  }\textbf {\bibinfo {volume} {147}},\ \bibinfo {pages} {224904} (\bibinfo
  {year} {2017})}\BibitemShut {NoStop}%
\bibitem [{\citenamefont {Chatterjee}(2000)}]{chatterjee:2000}%
  \BibitemOpen
  \bibfield  {author} {\bibinfo {author} {\bibfnamefont {A.~P.}\ \bibnamefont
  {Chatterjee}},\ }\href@noop {} {\bibfield  {journal} {\bibinfo  {journal} {J.
  Chem. Phys.}\ }\textbf {\bibinfo {volume} {113}},\ \bibinfo {pages} {9310}
  (\bibinfo {year} {2000})}\BibitemShut {NoStop}%
\bibitem [{\citenamefont {Wang}\ and\ \citenamefont
  {Chatterjee}(2003)}]{wang:2003}%
  \BibitemOpen
  \bibfield  {author} {\bibinfo {author} {\bibfnamefont {X.}~\bibnamefont
  {Wang}}\ and\ \bibinfo {author} {\bibfnamefont {A.}~\bibnamefont
  {Chatterjee}},\ }\href@noop {} {\bibfield  {journal} {\bibinfo  {journal} {J.
  Chem. Phys.}\ }\textbf {\bibinfo {volume} {118}},\ \bibinfo {pages} {10787}
  (\bibinfo {year} {2003})}\BibitemShut {NoStop}%
\bibitem [{\citenamefont {Yi}\ and\ \citenamefont {Sastry}(2004)}]{yi:2004}%
  \BibitemOpen
  \bibfield  {author} {\bibinfo {author} {\bibfnamefont {Y.-B.}\ \bibnamefont
  {Yi}}\ and\ \bibinfo {author} {\bibfnamefont {A.~M.}\ \bibnamefont
  {Sastry}},\ }\href@noop {} {\bibfield  {journal} {\bibinfo  {journal} {Proc.
  Roy. Soc. London A}\ }\textbf {\bibinfo {volume} {460}},\ \bibinfo {pages}
  {2353} (\bibinfo {year} {2004})}\BibitemShut {NoStop}%
\bibitem [{\citenamefont {Otten}\ and\ \citenamefont {van~der
  Schoot}(2009)}]{Otten2009}%
  \BibitemOpen
  \bibfield  {author} {\bibinfo {author} {\bibfnamefont {R.~H.~J.}\
  \bibnamefont {Otten}}\ and\ \bibinfo {author} {\bibfnamefont
  {P.}~\bibnamefont {van~der Schoot}},\ }\href {\doibase
  10.1103/PhysRevLett.103.225704} {\bibfield  {journal} {\bibinfo  {journal}
  {Phys. Rev. Lett.}\ }\textbf {\bibinfo {volume} {103}},\ \bibinfo {pages}
  {225704} (\bibinfo {year} {2009})}\BibitemShut {NoStop}%
\bibitem [{\citenamefont {Chatterjee}(2015)}]{chatterjee2015overview}%
  \BibitemOpen
  \bibfield  {author} {\bibinfo {author} {\bibfnamefont {A.~P.}\ \bibnamefont
  {Chatterjee}},\ }\href@noop {} {\bibfield  {journal} {\bibinfo  {journal}
  {Journal of Physics: Condensed Matter}\ }\textbf {\bibinfo {volume} {27}},\
  \bibinfo {pages} {375302} (\bibinfo {year} {2015})}\BibitemShut {NoStop}%
\bibitem [{\citenamefont {Kale}\ \emph {et~al.}(2015)\citenamefont {Kale},
  \citenamefont {Sabet}, \citenamefont {Jasiuk},\ and\ \citenamefont
  {Ostoja-Starzewski}}]{kale2015tunneling}%
  \BibitemOpen
  \bibfield  {author} {\bibinfo {author} {\bibfnamefont {S.}~\bibnamefont
  {Kale}}, \bibinfo {author} {\bibfnamefont {F.~A.}\ \bibnamefont {Sabet}},
  \bibinfo {author} {\bibfnamefont {I.}~\bibnamefont {Jasiuk}}, \ and\ \bibinfo
  {author} {\bibfnamefont {M.}~\bibnamefont {Ostoja-Starzewski}},\ }\href@noop
  {} {\bibfield  {journal} {\bibinfo  {journal} {Journal of Applied Physics}\
  }\textbf {\bibinfo {volume} {118}},\ \bibinfo {pages} {154306} (\bibinfo
  {year} {2015})}\BibitemShut {NoStop}%
\bibitem [{\citenamefont {Dixit}\ \emph {et~al.}(2016)\citenamefont {Dixit},
  \citenamefont {Meyer},\ and\ \citenamefont {Schilling}}]{Dixit16}%
  \BibitemOpen
  \bibfield  {author} {\bibinfo {author} {\bibfnamefont {M.}~\bibnamefont
  {Dixit}}, \bibinfo {author} {\bibfnamefont {H.}~\bibnamefont {Meyer}}, \ and\
  \bibinfo {author} {\bibfnamefont {T.}~\bibnamefont {Schilling}},\ }\href
  {\doibase 10.1103/PhysRevE.93.012116} {\bibfield  {journal} {\bibinfo
  {journal} {Phys. Rev. E}\ }\textbf {\bibinfo {volume} {93}},\ \bibinfo
  {pages} {012116} (\bibinfo {year} {2016})}\BibitemShut {NoStop}%
\bibitem [{\citenamefont {Schilling}\ \emph {et~al.}(2015)\citenamefont
  {Schilling}, \citenamefont {Miller},\ and\ \citenamefont {van~der
  Schoot}}]{Schilling15}%
  \BibitemOpen
  \bibfield  {author} {\bibinfo {author} {\bibfnamefont {T.}~\bibnamefont
  {Schilling}}, \bibinfo {author} {\bibfnamefont {M.~A.}\ \bibnamefont
  {Miller}}, \ and\ \bibinfo {author} {\bibfnamefont {P.}~\bibnamefont {van~der
  Schoot}},\ }\href {\doibase {10.1209/0295-5075/111/56004}} {\bibfield
  {journal} {\bibinfo  {journal} {{EPL}}\ }\textbf {\bibinfo {volume} {{111}}}
  (\bibinfo {year} {{2015}}),\ {10.1209/0295-5075/111/56004}}\BibitemShut
  {NoStop}%
\bibitem [{\citenamefont {Meyer}\ \emph {et~al.}(2015)\citenamefont {Meyer},
  \citenamefont {van~der Schoot},\ and\ \citenamefont {Schilling}}]{Meyer15}%
  \BibitemOpen
  \bibfield  {author} {\bibinfo {author} {\bibfnamefont {H.}~\bibnamefont
  {Meyer}}, \bibinfo {author} {\bibfnamefont {P.}~\bibnamefont {van~der
  Schoot}}, \ and\ \bibinfo {author} {\bibfnamefont {T.}~\bibnamefont
  {Schilling}},\ }\href {\doibase {10.1063/1.4926946}} {\bibfield  {journal}
  {\bibinfo  {journal} {{JOURNAL OF CHEMICAL PHYSICS}}\ }\textbf {\bibinfo
  {volume} {{143}}} (\bibinfo {year} {{2015}}),\
  {10.1063/1.4926946}}\BibitemShut {NoStop}%
\bibitem [{\citenamefont {Mecke}\ and\ \citenamefont
  {Seyfried}(2002)}]{mecke:2002}%
  \BibitemOpen
  \bibfield  {author} {\bibinfo {author} {\bibfnamefont {K.~R.}\ \bibnamefont
  {Mecke}}\ and\ \bibinfo {author} {\bibfnamefont {A.}~\bibnamefont
  {Seyfried}},\ }\href@noop {} {\bibfield  {journal} {\bibinfo  {journal}
  {Europhys. Lett.}\ }\textbf {\bibinfo {volume} {58}},\ \bibinfo {pages} {28}
  (\bibinfo {year} {2002})}\BibitemShut {NoStop}%
\bibitem [{\citenamefont {Pike}\ and\ \citenamefont
  {Seager}(1974)}]{pike:1974}%
  \BibitemOpen
  \bibfield  {author} {\bibinfo {author} {\bibfnamefont {G.~E.}\ \bibnamefont
  {Pike}}\ and\ \bibinfo {author} {\bibfnamefont {C.~H.}\ \bibnamefont
  {Seager}},\ }\href@noop {} {\bibfield  {journal} {\bibinfo  {journal} {Phys.
  Rev. B}\ }\textbf {\bibinfo {volume} {10}},\ \bibinfo {pages} {1421}
  (\bibinfo {year} {1974})}\BibitemShut {NoStop}%
\bibitem [{\citenamefont {Lee}\ and\ \citenamefont
  {Torquato}(1988)}]{lee:1988}%
  \BibitemOpen
  \bibfield  {author} {\bibinfo {author} {\bibfnamefont {S.~B.}\ \bibnamefont
  {Lee}}\ and\ \bibinfo {author} {\bibfnamefont {S.}~\bibnamefont {Torquato}},\
  }\href@noop {} {\bibfield  {journal} {\bibinfo  {journal} {J. Chem. Phys.}\
  }\textbf {\bibinfo {volume} {89}},\ \bibinfo {pages} {6427} (\bibinfo {year}
  {1988})}\BibitemShut {NoStop}%
\bibitem [{\citenamefont {Foygel}\ \emph {et~al.}(2005)\citenamefont {Foygel},
  \citenamefont {Morris}, \citenamefont {Anez}, \citenamefont {French},\ and\
  \citenamefont {Sobolev}}]{foygel:2005}%
  \BibitemOpen
  \bibfield  {author} {\bibinfo {author} {\bibfnamefont {M.}~\bibnamefont
  {Foygel}}, \bibinfo {author} {\bibfnamefont {R.~D.}\ \bibnamefont {Morris}},
  \bibinfo {author} {\bibfnamefont {D.}~\bibnamefont {Anez}}, \bibinfo {author}
  {\bibfnamefont {S.}~\bibnamefont {French}}, \ and\ \bibinfo {author}
  {\bibfnamefont {V.~L.}\ \bibnamefont {Sobolev}},\ }\href@noop {} {\bibfield
  {journal} {\bibinfo  {journal} {Phys. Rev. B}\ }\textbf {\bibinfo {volume}
  {71}},\ \bibinfo {pages} {104201} (\bibinfo {year} {2005})}\BibitemShut
  {NoStop}%
\bibitem [{\citenamefont {Rahatekar}\ \emph {et~al.}(2005)\citenamefont
  {Rahatekar}, \citenamefont {Hamm}, \citenamefont {Shaffer},\ and\
  \citenamefont {Elliott}}]{rahatekar2005mesoscale}%
  \BibitemOpen
  \bibfield  {author} {\bibinfo {author} {\bibfnamefont {S.~S.}\ \bibnamefont
  {Rahatekar}}, \bibinfo {author} {\bibfnamefont {M.}~\bibnamefont {Hamm}},
  \bibinfo {author} {\bibfnamefont {M.~S.}\ \bibnamefont {Shaffer}}, \ and\
  \bibinfo {author} {\bibfnamefont {J.~A.}\ \bibnamefont {Elliott}},\
  }\href@noop {} {\bibfield  {journal} {\bibinfo  {journal} {The Journal of
  chemical physics}\ }\textbf {\bibinfo {volume} {123}},\ \bibinfo {pages}
  {134702} (\bibinfo {year} {2005})}\BibitemShut {NoStop}%
\bibitem [{\citenamefont {Berhan}\ and\ \citenamefont
  {Sastry}(2007)}]{berhan:2007}%
  \BibitemOpen
  \bibfield  {author} {\bibinfo {author} {\bibfnamefont {L.}~\bibnamefont
  {Berhan}}\ and\ \bibinfo {author} {\bibfnamefont {A.~M.}\ \bibnamefont
  {Sastry}},\ }\href {\doibase 10.1103/PhysRevE.75.041120} {\bibfield
  {journal} {\bibinfo  {journal} {Phys. Rev. E}\ }\textbf {\bibinfo {volume}
  {75}},\ \bibinfo {pages} {041120} (\bibinfo {year} {2007})}\BibitemShut
  {NoStop}%
\bibitem [{\citenamefont {\ifmmode~\check{S}\else \v{S}\fi{}kvor}\ \emph
  {et~al.}(2007)\citenamefont {\ifmmode~\check{S}\else \v{S}\fi{}kvor},
  \citenamefont {Nezbeda}, \citenamefont {Brovchenko},\ and\ \citenamefont
  {Oleinikova}}]{Skvor07a}%
  \BibitemOpen
  \bibfield  {author} {\bibinfo {author} {\bibfnamefont {J.~c.~v.}\
  \bibnamefont {\ifmmode~\check{S}\else \v{S}\fi{}kvor}}, \bibinfo {author}
  {\bibfnamefont {I.}~\bibnamefont {Nezbeda}}, \bibinfo {author} {\bibfnamefont
  {I.}~\bibnamefont {Brovchenko}}, \ and\ \bibinfo {author} {\bibfnamefont
  {A.}~\bibnamefont {Oleinikova}},\ }\href {\doibase
  10.1103/PhysRevLett.99.127801} {\bibfield  {journal} {\bibinfo  {journal}
  {Phys. Rev. Lett.}\ }\textbf {\bibinfo {volume} {99}},\ \bibinfo {pages}
  {127801} (\bibinfo {year} {2007})}\BibitemShut {NoStop}%
\bibitem [{\citenamefont {Akagawa}\ and\ \citenamefont
  {Odagaki}(2007)}]{akagawa:2007}%
  \BibitemOpen
  \bibfield  {author} {\bibinfo {author} {\bibfnamefont {S.}~\bibnamefont
  {Akagawa}}\ and\ \bibinfo {author} {\bibfnamefont {T.}~\bibnamefont
  {Odagaki}},\ }\href@noop {} {\bibfield  {journal} {\bibinfo  {journal}
  {Physical Review E}\ }\textbf {\bibinfo {volume} {76}},\ \bibinfo {pages}
  {051402} (\bibinfo {year} {2007})}\BibitemShut {NoStop}%
\bibitem [{\citenamefont {Ambrosetti}\ \emph {et~al.}(2010)\citenamefont
  {Ambrosetti}, \citenamefont {Grimaldi}, \citenamefont {Balberg},
  \citenamefont {Maeder}, \citenamefont {Danani},\ and\ \citenamefont
  {Ryser}}]{ambrosetti:2010}%
  \BibitemOpen
  \bibfield  {author} {\bibinfo {author} {\bibfnamefont {G.}~\bibnamefont
  {Ambrosetti}}, \bibinfo {author} {\bibfnamefont {C.}~\bibnamefont
  {Grimaldi}}, \bibinfo {author} {\bibfnamefont {I.}~\bibnamefont {Balberg}},
  \bibinfo {author} {\bibfnamefont {T.}~\bibnamefont {Maeder}}, \bibinfo
  {author} {\bibfnamefont {A.}~\bibnamefont {Danani}}, \ and\ \bibinfo {author}
  {\bibfnamefont {P.}~\bibnamefont {Ryser}},\ }\href@noop {} {\bibfield
  {journal} {\bibinfo  {journal} {Phys. Rev. B}\ }\textbf {\bibinfo {volume}
  {81}},\ \bibinfo {pages} {155434} (\bibinfo {year} {2010})}\BibitemShut
  {NoStop}%
\bibitem [{\citenamefont {Schilling}\ \emph {et~al.}(2007)\citenamefont
  {Schilling}, \citenamefont {Jungblut},\ and\ \citenamefont
  {Miller}}]{schilling:2007}%
  \BibitemOpen
  \bibfield  {author} {\bibinfo {author} {\bibfnamefont {T.}~\bibnamefont
  {Schilling}}, \bibinfo {author} {\bibfnamefont {S.}~\bibnamefont {Jungblut}},
  \ and\ \bibinfo {author} {\bibfnamefont {M.~A.}\ \bibnamefont {Miller}},\
  }\href@noop {} {\bibfield  {journal} {\bibinfo  {journal} {Phys. Rev. Lett.}\
  }\textbf {\bibinfo {volume} {98}},\ \bibinfo {pages} {108303} (\bibinfo
  {year} {2007})}\BibitemShut {NoStop}%
\bibitem [{\citenamefont {Miller}(2009)}]{Miller09c}%
  \BibitemOpen
  \bibfield  {author} {\bibinfo {author} {\bibfnamefont {M.~A.}\ \bibnamefont
  {Miller}},\ }\href@noop {} {\bibfield  {journal} {\bibinfo  {journal} {J.
  Chem. Phys.}\ }\textbf {\bibinfo {volume} {131}},\ \bibinfo {pages} {066101}
  (\bibinfo {year} {2009})}\BibitemShut {NoStop}%
\bibitem [{\citenamefont {Mutiso}\ \emph {et~al.}(2012)\citenamefont {Mutiso},
  \citenamefont {Sherrott}, \citenamefont {Li},\ and\ \citenamefont
  {Winey}}]{mutiso:2012}%
  \BibitemOpen
  \bibfield  {author} {\bibinfo {author} {\bibfnamefont {R.~M.}\ \bibnamefont
  {Mutiso}}, \bibinfo {author} {\bibfnamefont {M.~C.}\ \bibnamefont
  {Sherrott}}, \bibinfo {author} {\bibfnamefont {J.}~\bibnamefont {Li}}, \ and\
  \bibinfo {author} {\bibfnamefont {K.~I.}\ \bibnamefont {Winey}},\ }\href@noop
  {} {\bibfield  {journal} {\bibinfo  {journal} {Phys. Rev. B}\ }\textbf
  {\bibinfo {volume} {86}},\ \bibinfo {pages} {214306} (\bibinfo {year}
  {2012})}\BibitemShut {NoStop}%
\bibitem [{\citenamefont {Mutiso}\ and\ \citenamefont
  {Winey}(2013)}]{Mutiso2013}%
  \BibitemOpen
  \bibfield  {author} {\bibinfo {author} {\bibfnamefont {R.~M.}\ \bibnamefont
  {Mutiso}}\ and\ \bibinfo {author} {\bibfnamefont {K.~I.}\ \bibnamefont
  {Winey}},\ }\href@noop {} {\bibfield  {journal} {\bibinfo  {journal}
  {Physical Review E - Statistical, Nonlinear, and Soft Matter Physics}\
  }\textbf {\bibinfo {volume} {88}} (\bibinfo {year} {2013})}\BibitemShut
  {NoStop}%
\bibitem [{\citenamefont {Coleman}\ \emph {et~al.}(1998)\citenamefont
  {Coleman}, \citenamefont {Curran}, \citenamefont {Dalton}, \citenamefont
  {Davey}, \citenamefont {McCarthy}, \citenamefont {Blau},\ and\ \citenamefont
  {Barklie}}]{coleman:1998}%
  \BibitemOpen
  \bibfield  {author} {\bibinfo {author} {\bibfnamefont {J.~N.}\ \bibnamefont
  {Coleman}}, \bibinfo {author} {\bibfnamefont {S.}~\bibnamefont {Curran}},
  \bibinfo {author} {\bibfnamefont {A.~B.}\ \bibnamefont {Dalton}}, \bibinfo
  {author} {\bibfnamefont {A.~P.}\ \bibnamefont {Davey}}, \bibinfo {author}
  {\bibfnamefont {B.}~\bibnamefont {McCarthy}}, \bibinfo {author}
  {\bibfnamefont {W.}~\bibnamefont {Blau}}, \ and\ \bibinfo {author}
  {\bibfnamefont {R.~C.}\ \bibnamefont {Barklie}},\ }\href {\doibase
  10.1103/PhysRevB.58.R7492} {\bibfield  {journal} {\bibinfo  {journal} {Phys.
  Rev. B}\ }\textbf {\bibinfo {volume} {58}},\ \bibinfo {pages} {R7492}
  (\bibinfo {year} {1998})}\BibitemShut {NoStop}%
\bibitem [{\citenamefont {Sandler~{\it et al}}(1999)}]{sandler:1999}%
  \BibitemOpen
  \bibfield  {author} {\bibinfo {author} {\bibfnamefont {J.}~\bibnamefont
  {Sandler~{\it et al}}},\ }\href@noop {} {\bibfield  {journal} {\bibinfo
  {journal} {Polymer}\ }\textbf {\bibinfo {volume} {40}},\ \bibinfo {pages}
  {5967} (\bibinfo {year} {1999})}\BibitemShut {NoStop}%
\bibitem [{\citenamefont {Barrau}\ \emph {et~al.}(2003)\citenamefont {Barrau},
  \citenamefont {Demont}, \citenamefont {Peigney}, \citenamefont {Laurent},\
  and\ \citenamefont {Lacabanne}}]{barrau:2003}%
  \BibitemOpen
  \bibfield  {author} {\bibinfo {author} {\bibfnamefont {S.}~\bibnamefont
  {Barrau}}, \bibinfo {author} {\bibfnamefont {P.}~\bibnamefont {Demont}},
  \bibinfo {author} {\bibfnamefont {A.}~\bibnamefont {Peigney}}, \bibinfo
  {author} {\bibfnamefont {C.}~\bibnamefont {Laurent}}, \ and\ \bibinfo
  {author} {\bibfnamefont {C.}~\bibnamefont {Lacabanne}},\ }\href@noop {}
  {\bibfield  {journal} {\bibinfo  {journal} {Macromolecules}\ }\textbf
  {\bibinfo {volume} {36}},\ \bibinfo {pages} {5187} (\bibinfo {year}
  {2003})}\BibitemShut {NoStop}%
\bibitem [{\citenamefont {Vigolo}\ \emph {et~al.}(2005)\citenamefont {Vigolo},
  \citenamefont {Coulon}, \citenamefont {Maugey}, \citenamefont {Zakri},\ and\
  \citenamefont {Poulin}}]{vigolo:2005}%
  \BibitemOpen
  \bibfield  {author} {\bibinfo {author} {\bibfnamefont {B.}~\bibnamefont
  {Vigolo}}, \bibinfo {author} {\bibfnamefont {C.}~\bibnamefont {Coulon}},
  \bibinfo {author} {\bibfnamefont {M.}~\bibnamefont {Maugey}}, \bibinfo
  {author} {\bibfnamefont {C.}~\bibnamefont {Zakri}}, \ and\ \bibinfo {author}
  {\bibfnamefont {P.}~\bibnamefont {Poulin}},\ }\href@noop {} {\bibfield
  {journal} {\bibinfo  {journal} {Science}\ }\textbf {\bibinfo {volume}
  {309}},\ \bibinfo {pages} {920} (\bibinfo {year} {2005})}\BibitemShut
  {NoStop}%
\bibitem [{\citenamefont {Lyons}\ \emph {et~al.}(2008)\citenamefont {Lyons},
  \citenamefont {De}, \citenamefont {Blighe}, \citenamefont {Nicolosi},
  \citenamefont {Pereira}, \citenamefont {Ferreira},\ and\ \citenamefont
  {Coleman}}]{lyons2008relationship}%
  \BibitemOpen
  \bibfield  {author} {\bibinfo {author} {\bibfnamefont {P.~E.}\ \bibnamefont
  {Lyons}}, \bibinfo {author} {\bibfnamefont {S.}~\bibnamefont {De}}, \bibinfo
  {author} {\bibfnamefont {F.}~\bibnamefont {Blighe}}, \bibinfo {author}
  {\bibfnamefont {V.}~\bibnamefont {Nicolosi}}, \bibinfo {author}
  {\bibfnamefont {L.~F.~C.}\ \bibnamefont {Pereira}}, \bibinfo {author}
  {\bibfnamefont {M.~S.}\ \bibnamefont {Ferreira}}, \ and\ \bibinfo {author}
  {\bibfnamefont {J.~N.}\ \bibnamefont {Coleman}},\ }\href@noop {} {\bibfield
  {journal} {\bibinfo  {journal} {Journal of Applied Physics}\ }\textbf
  {\bibinfo {volume} {104}},\ \bibinfo {pages} {044302} (\bibinfo {year}
  {2008})}\BibitemShut {NoStop}%
\bibitem [{\citenamefont {Nirmalraj}\ \emph {et~al.}(2012)\citenamefont
  {Nirmalraj}, \citenamefont {Bellew}, \citenamefont {Bell}, \citenamefont
  {Fairfield}, \citenamefont {McCarthy}, \citenamefont {O’Kelly},
  \citenamefont {Pereira}, \citenamefont {Sorel}, \citenamefont {Morosan},
  \citenamefont {Coleman} \emph {et~al.}}]{Nirmalraj:2012}%
  \BibitemOpen
  \bibfield  {author} {\bibinfo {author} {\bibfnamefont {P.~N.}\ \bibnamefont
  {Nirmalraj}}, \bibinfo {author} {\bibfnamefont {A.~T.}\ \bibnamefont
  {Bellew}}, \bibinfo {author} {\bibfnamefont {A.~P.}\ \bibnamefont {Bell}},
  \bibinfo {author} {\bibfnamefont {J.~A.}\ \bibnamefont {Fairfield}}, \bibinfo
  {author} {\bibfnamefont {E.~K.}\ \bibnamefont {McCarthy}}, \bibinfo {author}
  {\bibfnamefont {C.}~\bibnamefont {O’Kelly}}, \bibinfo {author}
  {\bibfnamefont {L.~F.}\ \bibnamefont {Pereira}}, \bibinfo {author}
  {\bibfnamefont {S.}~\bibnamefont {Sorel}}, \bibinfo {author} {\bibfnamefont
  {D.}~\bibnamefont {Morosan}}, \bibinfo {author} {\bibfnamefont {J.~N.}\
  \bibnamefont {Coleman}},  \emph {et~al.},\ }\href@noop {} {\bibfield
  {journal} {\bibinfo  {journal} {Nano letters}\ }\textbf {\bibinfo {volume}
  {12}},\ \bibinfo {pages} {5966} (\bibinfo {year} {2012})}\BibitemShut
  {NoStop}%
\bibitem [{\citenamefont {Cattin}\ and\ \citenamefont
  {Hubert}(2014)}]{Cattin14}%
  \BibitemOpen
  \bibfield  {author} {\bibinfo {author} {\bibfnamefont {C.}~\bibnamefont
  {Cattin}}\ and\ \bibinfo {author} {\bibfnamefont {P.}~\bibnamefont
  {Hubert}},\ }\href {\doibase 10.1021/am404808u} {\bibfield  {journal}
  {\bibinfo  {journal} {ACS APPLIED MATERIALS \& INTERFACES}\ }\textbf
  {\bibinfo {volume} {6}},\ \bibinfo {pages} {1804} (\bibinfo {year}
  {2014})}\BibitemShut {NoStop}%
\bibitem [{\citenamefont {Majidian}\ \emph {et~al.}(2017)\citenamefont
  {Majidian}, \citenamefont {Grimaldi}, \citenamefont {Forr{\'o}},\ and\
  \citenamefont {Magrez}}]{majidian2017role}%
  \BibitemOpen
  \bibfield  {author} {\bibinfo {author} {\bibfnamefont {M.}~\bibnamefont
  {Majidian}}, \bibinfo {author} {\bibfnamefont {C.}~\bibnamefont {Grimaldi}},
  \bibinfo {author} {\bibfnamefont {L.}~\bibnamefont {Forr{\'o}}}, \ and\
  \bibinfo {author} {\bibfnamefont {A.}~\bibnamefont {Magrez}},\ }\href@noop {}
  {\bibfield  {journal} {\bibinfo  {journal} {Scientific reports}\ }\textbf
  {\bibinfo {volume} {7}},\ \bibinfo {pages} {12553} (\bibinfo {year}
  {2017})}\BibitemShut {NoStop}%
\bibitem [{\citenamefont {Grossiord}\ \emph {et~al.}(2008)\citenamefont
  {Grossiord}, \citenamefont {Kivit}, \citenamefont {Loos}, \citenamefont
  {Meuldijk}, \citenamefont {Kyrylyuk}, \citenamefont {van~der Schoot},\ and\
  \citenamefont {Koning}}]{Grossiord2008}%
  \BibitemOpen
  \bibfield  {author} {\bibinfo {author} {\bibfnamefont {N.}~\bibnamefont
  {Grossiord}}, \bibinfo {author} {\bibfnamefont {P.~J.}\ \bibnamefont
  {Kivit}}, \bibinfo {author} {\bibfnamefont {J.}~\bibnamefont {Loos}},
  \bibinfo {author} {\bibfnamefont {J.}~\bibnamefont {Meuldijk}}, \bibinfo
  {author} {\bibfnamefont {A.~V.}\ \bibnamefont {Kyrylyuk}}, \bibinfo {author}
  {\bibfnamefont {P.}~\bibnamefont {van~der Schoot}}, \ and\ \bibinfo {author}
  {\bibfnamefont {C.~E.}\ \bibnamefont {Koning}},\ }\href {\doibase
  10.1016/j.polymer.2008.04.033} {\enquote {\bibinfo {title} {{On the influence
  of the processing conditions on the performance of electrically conductive
  carbon nanotube/polymer nanocomposites}},}\ } (\bibinfo {year}
  {2008})\BibitemShut {NoStop}%
\bibitem [{\citenamefont {Kyrylyuk}\ \emph {et~al.}(2011)\citenamefont
  {Kyrylyuk}, \citenamefont {Hermant}, \citenamefont {Schilling}, \citenamefont
  {Klumperman}, \citenamefont {Koning},\ and\ \citenamefont {van~der
  Schoot}}]{Kyrylyuk2011}%
  \BibitemOpen
  \bibfield  {author} {\bibinfo {author} {\bibfnamefont {A.~V.}\ \bibnamefont
  {Kyrylyuk}}, \bibinfo {author} {\bibfnamefont {M.~C.}\ \bibnamefont
  {Hermant}}, \bibinfo {author} {\bibfnamefont {T.}~\bibnamefont {Schilling}},
  \bibinfo {author} {\bibfnamefont {B.}~\bibnamefont {Klumperman}}, \bibinfo
  {author} {\bibfnamefont {C.~E.}\ \bibnamefont {Koning}}, \ and\ \bibinfo
  {author} {\bibfnamefont {P.}~\bibnamefont {van~der Schoot}},\ }\href
  {\doibase 10.1038/nnano.2011.40} {\bibfield  {journal} {\bibinfo  {journal}
  {Nature nanotechnology}\ }\textbf {\bibinfo {volume} {6}},\ \bibinfo {pages}
  {364} (\bibinfo {year} {2011})}\BibitemShut {NoStop}%
\bibitem [{\citenamefont {Kumar}\ and\ \citenamefont
  {Rawal}(2016)}]{kumar2016tuning}%
  \BibitemOpen
  \bibfield  {author} {\bibinfo {author} {\bibfnamefont {V.}~\bibnamefont
  {Kumar}}\ and\ \bibinfo {author} {\bibfnamefont {A.}~\bibnamefont {Rawal}},\
  }\href@noop {} {\bibfield  {journal} {\bibinfo  {journal} {Polymer}\ }\textbf
  {\bibinfo {volume} {97}},\ \bibinfo {pages} {295} (\bibinfo {year}
  {2016})}\BibitemShut {NoStop}%
\bibitem [{\citenamefont {Onsager}(1949)}]{onsager1949effects}%
  \BibitemOpen
  \bibfield  {author} {\bibinfo {author} {\bibfnamefont {L.}~\bibnamefont
  {Onsager}},\ }\href@noop {} {\bibfield  {journal} {\bibinfo  {journal}
  {Annals of the New York Academy of Sciences}\ }\textbf {\bibinfo {volume}
  {51}},\ \bibinfo {pages} {627} (\bibinfo {year} {1949})}\BibitemShut
  {NoStop}%
\bibitem [{\citenamefont {de~Gennes}\ and\ \citenamefont
  {Prost}(1995)}]{p1995physics}%
  \BibitemOpen
  \bibfield  {author} {\bibinfo {author} {\bibfnamefont {P.~G.}\ \bibnamefont
  {de~Gennes}}\ and\ \bibinfo {author} {\bibfnamefont {J.}~\bibnamefont
  {Prost}},\ }\href {https://books.google.de/books?id=0Nw-dzWz5agC} {\emph
  {\bibinfo {title} {The Physics of Liquid Crystals}}},\ International Series
  of Monographs on Physics, 83\ (\bibinfo  {publisher} {Clarendon Press},\
  \bibinfo {year} {1995})\BibitemShut {NoStop}%
\bibitem [{\citenamefont {Ackermann}\ \emph {et~al.}(2016)\citenamefont
  {Ackermann}, \citenamefont {Neuhaus},\ and\ \citenamefont
  {Roth}}]{ackermann2016effect}%
  \BibitemOpen
  \bibfield  {author} {\bibinfo {author} {\bibfnamefont {T.}~\bibnamefont
  {Ackermann}}, \bibinfo {author} {\bibfnamefont {R.}~\bibnamefont {Neuhaus}},
  \ and\ \bibinfo {author} {\bibfnamefont {S.}~\bibnamefont {Roth}},\
  }\href@noop {} {\bibfield  {journal} {\bibinfo  {journal} {Scientific
  reports}\ }\textbf {\bibinfo {volume} {6}},\ \bibinfo {pages} {34289}
  (\bibinfo {year} {2016})}\BibitemShut {NoStop}%
\bibitem [{\citenamefont {Wan}\ \emph {et~al.}(2017)\citenamefont {Wan},
  \citenamefont {Song}, \citenamefont {Yang}, \citenamefont {Kirsch},
  \citenamefont {Jia}, \citenamefont {Xu}, \citenamefont {Dai}, \citenamefont
  {Zhu}, \citenamefont {Xu}, \citenamefont {Chen} \emph
  {et~al.}}]{wan2017highly}%
  \BibitemOpen
  \bibfield  {author} {\bibinfo {author} {\bibfnamefont {J.}~\bibnamefont
  {Wan}}, \bibinfo {author} {\bibfnamefont {J.}~\bibnamefont {Song}}, \bibinfo
  {author} {\bibfnamefont {Z.}~\bibnamefont {Yang}}, \bibinfo {author}
  {\bibfnamefont {D.}~\bibnamefont {Kirsch}}, \bibinfo {author} {\bibfnamefont
  {C.}~\bibnamefont {Jia}}, \bibinfo {author} {\bibfnamefont {R.}~\bibnamefont
  {Xu}}, \bibinfo {author} {\bibfnamefont {J.}~\bibnamefont {Dai}}, \bibinfo
  {author} {\bibfnamefont {M.}~\bibnamefont {Zhu}}, \bibinfo {author}
  {\bibfnamefont {L.}~\bibnamefont {Xu}}, \bibinfo {author} {\bibfnamefont
  {C.}~\bibnamefont {Chen}},  \emph {et~al.},\ }\href@noop {} {\bibfield
  {journal} {\bibinfo  {journal} {Advanced Materials}\ }\textbf {\bibinfo
  {volume} {29}},\ \bibinfo {pages} {1703331} (\bibinfo {year}
  {2017})}\BibitemShut {NoStop}%
\bibitem [{\citenamefont {Wang}\ \emph {et~al.}(2008)\citenamefont {Wang},
  \citenamefont {Dai}, \citenamefont {Li}, \citenamefont {Wei},\ and\
  \citenamefont {Jiang}}]{wang2008effects}%
  \BibitemOpen
  \bibfield  {author} {\bibinfo {author} {\bibfnamefont {Q.}~\bibnamefont
  {Wang}}, \bibinfo {author} {\bibfnamefont {J.}~\bibnamefont {Dai}}, \bibinfo
  {author} {\bibfnamefont {W.}~\bibnamefont {Li}}, \bibinfo {author}
  {\bibfnamefont {Z.}~\bibnamefont {Wei}}, \ and\ \bibinfo {author}
  {\bibfnamefont {J.}~\bibnamefont {Jiang}},\ }\href@noop {} {\bibfield
  {journal} {\bibinfo  {journal} {Composites science and technology}\ }\textbf
  {\bibinfo {volume} {68}},\ \bibinfo {pages} {1644} (\bibinfo {year}
  {2008})}\BibitemShut {NoStop}%
\bibitem [{\citenamefont {Choi}\ \emph {et~al.}(2003)\citenamefont {Choi},
  \citenamefont {Brooks}, \citenamefont {Eaton}, \citenamefont {Al-Haik},
  \citenamefont {Hussaini}, \citenamefont {Garmestani}, \citenamefont {Li},\
  and\ \citenamefont {Dahmen}}]{choi2003enhancement}%
  \BibitemOpen
  \bibfield  {author} {\bibinfo {author} {\bibfnamefont {E.}~\bibnamefont
  {Choi}}, \bibinfo {author} {\bibfnamefont {J.}~\bibnamefont {Brooks}},
  \bibinfo {author} {\bibfnamefont {D.}~\bibnamefont {Eaton}}, \bibinfo
  {author} {\bibfnamefont {M.}~\bibnamefont {Al-Haik}}, \bibinfo {author}
  {\bibfnamefont {M.}~\bibnamefont {Hussaini}}, \bibinfo {author}
  {\bibfnamefont {H.}~\bibnamefont {Garmestani}}, \bibinfo {author}
  {\bibfnamefont {D.}~\bibnamefont {Li}}, \ and\ \bibinfo {author}
  {\bibfnamefont {K.}~\bibnamefont {Dahmen}},\ }\href@noop {} {\bibfield
  {journal} {\bibinfo  {journal} {Journal of Applied physics}\ }\textbf
  {\bibinfo {volume} {94}},\ \bibinfo {pages} {6034} (\bibinfo {year}
  {2003})}\BibitemShut {NoStop}%
\bibitem [{\citenamefont {Du}\ \emph {et~al.}(2005)\citenamefont {Du},
  \citenamefont {Fischer},\ and\ \citenamefont {Winey}}]{du2005effect}%
  \BibitemOpen
  \bibfield  {author} {\bibinfo {author} {\bibfnamefont {F.}~\bibnamefont
  {Du}}, \bibinfo {author} {\bibfnamefont {J.~E.}\ \bibnamefont {Fischer}}, \
  and\ \bibinfo {author} {\bibfnamefont {K.~I.}\ \bibnamefont {Winey}},\
  }\href@noop {} {\bibfield  {journal} {\bibinfo  {journal} {Physical Review
  B}\ }\textbf {\bibinfo {volume} {72}},\ \bibinfo {pages} {121404} (\bibinfo
  {year} {2005})}\BibitemShut {NoStop}%
\bibitem [{\citenamefont {Finner}\ \emph {et~al.}(2018)\citenamefont {Finner},
  \citenamefont {Kotsev}, \citenamefont {Miller},\ and\ \citenamefont {van~der
  Schoot}}]{finner2018continuum}%
  \BibitemOpen
  \bibfield  {author} {\bibinfo {author} {\bibfnamefont {S.~P.}\ \bibnamefont
  {Finner}}, \bibinfo {author} {\bibfnamefont {M.~I.}\ \bibnamefont {Kotsev}},
  \bibinfo {author} {\bibfnamefont {M.~A.}\ \bibnamefont {Miller}}, \ and\
  \bibinfo {author} {\bibfnamefont {P.}~\bibnamefont {van~der Schoot}},\
  }\href@noop {} {\bibfield  {journal} {\bibinfo  {journal} {The Journal of
  chemical physics}\ }\textbf {\bibinfo {volume} {148}},\ \bibinfo {pages}
  {034903} (\bibinfo {year} {2018})}\BibitemShut {NoStop}%
\bibitem [{\citenamefont {Kale}\ \emph {et~al.}(2016)\citenamefont {Kale},
  \citenamefont {Sabet}, \citenamefont {Jasiuk},\ and\ \citenamefont
  {Ostoja-Starzewski}}]{kale2016effect}%
  \BibitemOpen
  \bibfield  {author} {\bibinfo {author} {\bibfnamefont {S.}~\bibnamefont
  {Kale}}, \bibinfo {author} {\bibfnamefont {F.~A.}\ \bibnamefont {Sabet}},
  \bibinfo {author} {\bibfnamefont {I.}~\bibnamefont {Jasiuk}}, \ and\ \bibinfo
  {author} {\bibfnamefont {M.}~\bibnamefont {Ostoja-Starzewski}},\ }\href@noop
  {} {\bibfield  {journal} {\bibinfo  {journal} {Journal of Applied Physics}\
  }\textbf {\bibinfo {volume} {120}},\ \bibinfo {pages} {045105} (\bibinfo
  {year} {2016})}\BibitemShut {NoStop}%
\bibitem [{\citenamefont {Zheng}\ \emph {et~al.}(2005)\citenamefont {Zheng},
  \citenamefont {Forest}, \citenamefont {Lipton}, \citenamefont {Zhou},\ and\
  \citenamefont {Wang}}]{zheng:2005}%
  \BibitemOpen
  \bibfield  {author} {\bibinfo {author} {\bibfnamefont {X.}~\bibnamefont
  {Zheng}}, \bibinfo {author} {\bibfnamefont {M.~G.}\ \bibnamefont {Forest}},
  \bibinfo {author} {\bibfnamefont {R.}~\bibnamefont {Lipton}}, \bibinfo
  {author} {\bibfnamefont {R.}~\bibnamefont {Zhou}}, \ and\ \bibinfo {author}
  {\bibfnamefont {Q.}~\bibnamefont {Wang}},\ }\href@noop {} {\bibfield
  {journal} {\bibinfo  {journal} {Advanced Functional Materials}\ }\textbf
  {\bibinfo {volume} {15}},\ \bibinfo {pages} {627} (\bibinfo {year}
  {2005})}\BibitemShut {NoStop}%
\bibitem [{\citenamefont {Otten}\ and\ \citenamefont {van~der
  Schoot}(2012)}]{Otten12}%
  \BibitemOpen
  \bibfield  {author} {\bibinfo {author} {\bibfnamefont {R.~H.~J.}\
  \bibnamefont {Otten}}\ and\ \bibinfo {author} {\bibfnamefont
  {P.}~\bibnamefont {van~der Schoot}},\ }\href {\doibase
  {10.1103/PhysRevLett.108.088301}} {\bibfield  {journal} {\bibinfo  {journal}
  {{PHYSICAL REVIEW LETTERS}}\ }\textbf {\bibinfo {volume} {{108}}} (\bibinfo
  {year} {{2012}}),\ {10.1103/PhysRevLett.108.088301}}\BibitemShut {NoStop}%
\bibitem [{\citenamefont {Lebovka}\ \emph {et~al.}(2018)\citenamefont
  {Lebovka}, \citenamefont {Tarasevich}, \citenamefont {Vygornitskii},
  \citenamefont {Eserkepov},\ and\ \citenamefont
  {Akhunzhanov}}]{lebovka2018anisotropy}%
  \BibitemOpen
  \bibfield  {author} {\bibinfo {author} {\bibfnamefont {N.~I.}\ \bibnamefont
  {Lebovka}}, \bibinfo {author} {\bibfnamefont {Y.~Y.}\ \bibnamefont
  {Tarasevich}}, \bibinfo {author} {\bibfnamefont {N.~V.}\ \bibnamefont
  {Vygornitskii}}, \bibinfo {author} {\bibfnamefont {A.~V.}\ \bibnamefont
  {Eserkepov}}, \ and\ \bibinfo {author} {\bibfnamefont {R.~K.}\ \bibnamefont
  {Akhunzhanov}},\ }\href@noop {} {\bibfield  {journal} {\bibinfo  {journal}
  {Physical Review E}\ }\textbf {\bibinfo {volume} {98}},\ \bibinfo {pages}
  {012104} (\bibinfo {year} {2018})}\BibitemShut {NoStop}%
\bibitem [{Note1()}]{Note1}%
  \BibitemOpen
  \bibinfo {note} {Where the director of the nematic phase was roughly parallel
  to the elongated z-direction}\BibitemShut {NoStop}%
\bibitem [{\citenamefont {Sherman}\ \emph {et~al.}(1983)\citenamefont
  {Sherman}, \citenamefont {Middleman},\ and\ \citenamefont
  {Jacobs}}]{sherman1983electron}%
  \BibitemOpen
  \bibfield  {author} {\bibinfo {author} {\bibfnamefont {R.}~\bibnamefont
  {Sherman}}, \bibinfo {author} {\bibfnamefont {L.}~\bibnamefont {Middleman}},
  \ and\ \bibinfo {author} {\bibfnamefont {S.}~\bibnamefont {Jacobs}},\
  }\href@noop {} {\bibfield  {journal} {\bibinfo  {journal} {Polymer
  Engineering \& Science}\ }\textbf {\bibinfo {volume} {23}},\ \bibinfo {pages}
  {36} (\bibinfo {year} {1983})}\BibitemShut {NoStop}%
\bibitem [{\citenamefont {Nigro}\ and\ \citenamefont
  {Grimaldi}(2014)}]{nigro2014impact}%
  \BibitemOpen
  \bibfield  {author} {\bibinfo {author} {\bibfnamefont {B.}~\bibnamefont
  {Nigro}}\ and\ \bibinfo {author} {\bibfnamefont {C.}~\bibnamefont
  {Grimaldi}},\ }\href {\doibase 10.1103/PhysRevB.90.094202} {\bibfield
  {journal} {\bibinfo  {journal} {Phys. Rev. B}\ }\textbf {\bibinfo {volume}
  {90}},\ \bibinfo {pages} {094202} (\bibinfo {year} {2014})}\BibitemShut
  {NoStop}%
\bibitem [{\citenamefont {Chatterjee}\ and\ \citenamefont
  {Grimaldi}(2015)}]{chatterjee2015tunneling}%
  \BibitemOpen
  \bibfield  {author} {\bibinfo {author} {\bibfnamefont {A.~P.}\ \bibnamefont
  {Chatterjee}}\ and\ \bibinfo {author} {\bibfnamefont {C.}~\bibnamefont
  {Grimaldi}},\ }\href@noop {} {\bibfield  {journal} {\bibinfo  {journal}
  {Journal of Physics: Condensed Matter}\ }\textbf {\bibinfo {volume} {27}},\
  \bibinfo {pages} {145302} (\bibinfo {year} {2015})}\BibitemShut {NoStop}%
\bibitem [{\citenamefont {Margraf}\ \emph {et~al.}(2017)\citenamefont
  {Margraf}, \citenamefont {Jahn}, \citenamefont {Flucke}, \citenamefont
  {Jacob}, \citenamefont {Habchi}, \citenamefont {Ishikawa}, \citenamefont
  {Krishna}, \citenamefont {Brinson}, \citenamefont {Parruitte}, \citenamefont
  {Roucaries} \emph {et~al.}}]{margraf2017qucs}%
  \BibitemOpen
  \bibfield  {author} {\bibinfo {author} {\bibfnamefont {M.}~\bibnamefont
  {Margraf}}, \bibinfo {author} {\bibfnamefont {S.}~\bibnamefont {Jahn}},
  \bibinfo {author} {\bibfnamefont {J.}~\bibnamefont {Flucke}}, \bibinfo
  {author} {\bibfnamefont {R.}~\bibnamefont {Jacob}}, \bibinfo {author}
  {\bibfnamefont {V.}~\bibnamefont {Habchi}}, \bibinfo {author} {\bibfnamefont
  {T.}~\bibnamefont {Ishikawa}}, \bibinfo {author} {\bibfnamefont {A.~G.}\
  \bibnamefont {Krishna}}, \bibinfo {author} {\bibfnamefont {M.}~\bibnamefont
  {Brinson}}, \bibinfo {author} {\bibfnamefont {H.}~\bibnamefont {Parruitte}},
  \bibinfo {author} {\bibfnamefont {B.}~\bibnamefont {Roucaries}},  \emph
  {et~al.},\ }\href {http://qucs.sourceforge.net/index.html} {\enquote
  {\bibinfo {title} {Qucs (quite universal circuit simulator), version
  0.0.19},}\ } (\bibinfo {year} {2017})\BibitemShut {NoStop}%
\bibitem [{\citenamefont {Middendorf}(1956)}]{middendorf1956analysis}%
  \BibitemOpen
  \bibfield  {author} {\bibinfo {author} {\bibfnamefont {W.~H.}\ \bibnamefont
  {Middendorf}},\ }\href@noop {} {\emph {\bibinfo {title} {Analysis of electric
  circuits}}}\ (\bibinfo  {publisher} {Wiley},\ \bibinfo {year}
  {1956})\BibitemShut {NoStop}%
\bibitem [{Note2()}]{Note2}%
  \BibitemOpen
  \bibinfo {note} {On a related note, for an analysis based on the effective
  medium approximation (EMA) and critical path approximation (CPA) of the
  dependence of $\sigma $ on the interplay of volume fraction and orientational
  ordering, we refer the reader to the work of A. P. Chatterjee and C.
  Grimaldi~\cite {chatterjee2015tunneling}.}\BibitemShut {Stop}%
\bibitem [{Note3()}]{Note3}%
  \BibitemOpen
  \bibinfo {note} {More precisely, first an effective expression for the
  electrical conductivity tensor is derived, then the largest relative
  enhancement is defined in terms of the maximum eigenvalue of the effective
  conductivity tensor after the inclusion of conductive filler rods. For a
  clear account of all details, we refer the reader to the complete work of
  Zheng et al.~\cite {zheng:2005}.}\BibitemShut {Stop}%
\end{thebibliography}


\end{document}